\documentstyle[epsfig]{mn2e}

\def\gtsima{$\; \buildrel > \over \sim \;$}
\def\ltsima{$\; \buildrel < \over \sim \;$}
\def\gtrsim{\lower.5ex\hbox{\gtsima}}
\def\lesssim{\lower.5ex\hbox{\ltsima}}

\begin{document}

\title[A minor merger scenario for HLX-1 - II. Photometry]{A minor merger scenario for the ultraluminous X-ray source ESO 243-49 HLX-1 - II. Constraints from photometry}
\author[Mapelli et al.]
{M. Mapelli$^{1}$, F. Annibali$^{2}$, L. Zampieri$^{1}$, R. Soria$^{3}$
\\
$^1$INAF-Osservatorio Astronomico di Padova, Vicolo dell'Osservatorio 5, I--35122, Padova, Italy, {\tt michela.mapelli@oapd.inaf.it}\\
$^2$INAF-Osservatorio Astronomico di Bologna, Via Ranzani 1, I--40127 Bologna, Italy \\
$^3$Curtin Institute of Radio Astronomy, Curtin University, 1 Turner Avenue, Bentley, WA 6102, Australia \\
}
\maketitle \vspace {7cm }

\begin{abstract}
The point-like X-ray source HLX-1, close to the S0 galaxy ESO 243-49, is the brightest known ultraluminous X-ray source and one the strongest intermediate-mass black hole candidates. We argue that the counterpart of  HLX-1 may be the nucleus of a satellite galaxy, undergoing minor merger with the S0 galaxy. We investigate this scenario by running a set of $N-$body/smoothed particle hydrodynamics simulations of the minor merger between a S0 galaxy and a gas-rich bulgy satellite galaxy,  and by comparing the results with the available photometric {\it Hubble Space Telescope} ({\it HST}) data of  ESO 243-49 and of the HLX-1 counterpart.
In particular, we derive synthetic surface brightness profiles for the simulated counterpart of HLX-1 in six {\it HST} filters, ranging from far ultraviolet (FUV) to infrared wavelengths. Such synthetic profiles include a contribution from the stellar population associated with the simulated disrupted satellite and a contribution from an irradiated disc model. These are in reasonable agreement with the observed   surface brightness profiles of the HLX-1 counterpart, provided that  the merger is at sufficiently late stage ($\gtrsim{}2.5$ Gyr since the first pericentre passage). 
The main difference between models and observations is in the FUV band, where the {\it HST} image shows a fuzzy and extended emission.

We show that the spectral energy distribution of the bulge of ESO~243-49 cannot be explained with a single old stellar population, but requires the existence of a younger stellar component. This is in good agreement with the star formation history derived from our $N-$body simulations, and is a further hint for the minor merger scenario.
\end{abstract}

%

\section{Introduction}~\label{sec:intro}
The point-like X-ray source 2XMM~J011028.1$-$460421 (hereafter HLX-1, Farrell et al. 2009; Godet et al. 2009) is the brightest known ultraluminous X-ray source (ULX, see Feng \&{} Soria 2011 for a review) and one of the strongest intermediate-mass black hole (IMBH) candidates (see van der Marel 2004 for a review).  The most recent estimate for the black hole (BH) mass is $\sim{}2\times{}10^4$ M$_\odot{}$ (Godet et al. 2012; see also Davis et al. 2011 and Servillat et al. 2011 for previous estimates). HLX-1 is located in the outskirts of the S0/a galaxy ESO~243-49 (luminosity distance $\sim{}96$ Mpc), $\sim{}0.8$ kpc out of the plane and $\sim{}3.3$ kpc away from the nucleus. The galaxy ESO~243-49 is a member of the cluster Abell~2877 (e.g. Malumuth et al. 1992).

The X-ray variability of HLX-1, with a semi-regular period of $\sim{}370$ days, may be connected with the orbital period of the companion star (Lasota et al. 2011).  HLX-1 has an optical counterpart (Wiersema et al. 2010; Soria et al. 2010, 2012, hereafter S10, S12, respectively; Farrell et al. 2012, hereafter F12), detected in various bands, from near infrared to far ultraviolet (UV). The  vicinity of HLX-1 with ESO~243-49 is confirmed by the redshift of the observed H$\alpha{}$ emission line of the counterpart (Wiersema et al. 2010; Soria \&{} Hau 2012; Soria, Hau \&{} Pakull 2013).

The nature of the HLX-1 counterpart is uncertain. Fits to the {\it Hubble Space Telescope} ({\it HST}) data (F12) indicate a total stellar mass of $4-6\times{}10^6$ M$_\odot{}$.  Considering the {\it HST} data alone, both a very young ($\sim{}10$ Myr) and a very old ($\sim{}10-13$ Gyr) age are possible, while an intermediate-age solution is excluded. It is also possible that the counterpart of HLX-1 has a multiple population, whose bulk is composed of old stars, plus a secondary young component (e.g. Mapelli et al. 2012, hereafter paper~I). Fitting the data with a very young stellar population requires minimal reprocessing from the disc, whereas the old population explanation predicts that most of the blue/UV emission  is due to disc irradiation (F12). From the analysis of Very Large Telescope (VLT) data in the U, B, V, R and I filters, S12 argue that the optical/UV component  is variable and therefore dominated by disc irradiation, and exclude the scenario of a $\gg{}10^4$ M$_\odot{}$ young star cluster (SC). Farrell et al.  argue that the apparent difference between {\it HST} and VLT fluxes is due to an imperfect subtraction of the diffuse galaxy emission in the VLT images, which have a much larger point spread function (S. Farrell, private communication). If the variability is confirmed, the two possible scenarios for the counterpart are an old $\sim{}10^6$ M$_\odot{}$ SC and a young $\approx{}10^4$ M$_\odot{}$  SC. The latter scenario is unlikely, as very massive ($>10^2$ M$_\odot{}$) BHs cannot form in such small SCs, according to theoretical models (e.g., Portegies Zwart \&{} McMillan 2002). Furthermore, most of young SCs are a disc population (e.g., Portegies Zwart, McMillan \&{} Gieles 2010, for a recent review), while the counterpart of HLX-1 is significantly offset with respect to the disc of ESO~243-49.

Similarly, it is not easy to explain how  a  $\sim{}10^4$ M$_\odot{}$ IMBH can form in a $\sim{}10^6$ M$_\odot{}$ SC. Repeated mergers between stellar BHs and other compact objects and/or stars were proposed to produce IMBHs in globular clusters (e.g. Miller $\&$ Hamilton 2002), while runaway collapse of massive stars can occur in young massive SCs (e.g., Portegies Zwart \&{} McMillan 2002). 

An alternative scenario predicts that the IMBH is associated with the nucleus of a satellite galaxy which is being disrupted in a minor merger with  ESO~243-49 (Bellovary et al. 2010; S10; Webb et al. 2010; paper~I). In this case, the IMBH would belong to the low-mass tail of the distribution of super massive BHs (SMBHs), located at the centre of galaxies. There is increasing evidence of (both bulgy and bulgeless) galaxies hosting at their centre SMBHs with mass $\lesssim{}10^5$ M$_\odot{}$ (e.g. Filippenko \&{} Sargent 1989; Filippenko \&{} Ho 2003; Barth et al. 2004; Greene \&{} Ho 2004, 2007a, 2007b; Satyapal et al. 2007, 2008, 2009; Dewangen et al. 2008; Shields et al. 2008; Barth et al. 2009; Desroches \&{} Ho 2009; Gliozzi et al. 2009; Jiang et al. 2011a, 2011b; Secrest et al. 2012). 

The hypothesis that  ESO~243-49 recently underwent a minor merger is suggested by various features, such as the presence of prominent dust lanes around its nucleus  (Finkelman et al. 2010; Kaviraj et al. 2012; Shabala et al. 2012) and the evidence of UV emission centred on its bulge (S10), indicating ongoing star formation (SF, Kaviraj et al. 2009, 2011; paper~I).

In paper~I, we presented the results of two preliminary $N-$body simulations, showing that the minor merger scenario is viable to explain the point-like appearance of the HLX-1 counterpart, the relative velocity between it and the centre of the S0, and the observed SF rate (SFR).

In the current paper, we investigate the minor-merger scenario in more detail, by analyzing a set of six simulations (among which one very high-resolution run) and by comparing the simulation outputs with the {\it HST} data. In particular, we compare the surface brightness profiles estimated from the simulations  with the observations, from infrared to UV bands. We also extract new accurate estimates of the spectral energy distribution (SED) of ESO~243-49 from the data and compare them with the simulated SF history.
  The paper is organized as follows. In Section~\ref{sec:method}, we describe the method adopted for data analysis and for the $N-$body simulations. In Section~\ref{sec:hlx1}, we discuss the (UV, optical and infrared) properties of the HLX-1 counterpart. In particular, synthetic surface brightness profiles of the HLX-1 counterpart are derived from the models and compared with the observations. In Section~\ref{sec:SEDbulge}, we present the SF histories predicted by the simulations and we compare them with the observational SED of the bulge of ESO~243-49. In Section~\ref{sec:discussion} we discuss the arguments in favour and against the minor-merger scenario. The main conclusions are summarized in Section~\ref{sec:conclude}.
\section{Method}~\label{sec:method}
\subsection{Data analysis}~\label{sec:data}
\begin{table*}
\begin{center}
\caption{{\it HST} photometry of the HLX-1 counterpart.}
 \leavevmode
\begin{tabular}[!h]{cccccc}
\hline
Band & Filter & $\lambda{}_{\rm pivot}$ & $m(<0.4'')$ & $m$                         & PSF FWHM\\
     &        & (\AA{})   & (Vegamag)   &  (ABmag, Vegamag)      & (arcsec) \\
\hline
FUV & F140LP & 1527    & $22.21 \pm 0.03$ & $24.33$, $22.16 \pm 0.03$ & 0.11\\
NUV & F300X  & 2829.8  & $22.80 \pm 0.05$ & $24.04$, $22.62 \pm 0.05$ & 0.08\\
C   & F390W  & 3904.6  & $24.04 \pm 0.05$ & $24.13$, $23.92 \pm 0.05$ & 0.09\\
V   & F555W  & 5309.8  & $24.11 \pm 0.05$ & $23.98$, $24.01 \pm 0.05$ & 0.09\\
I   & F775W  & 7733.6  & $23.64 \pm 0.15$ & $23.91$, $23.53 \pm 0.15$ & 0.09\\
H   & F160W  & 15405.2 & $23.49 \pm 0.26$ & $24.55$, $23.30 \pm 0.26$ & 0.3 \\
\noalign{\vspace{0.1cm}}
\hline
\end{tabular}
\begin{flushleft}
\footnotesize{$m(<0.4'')$ is the magnitude within  a radius of 0.4 arcsec, derived as described in Section~\ref{sec:data}.  $m$ is the total magnitude corrected for an infinite aperture assuming that the source is point-like (see description in Section~\ref{sec:data}). We report $m$ in both Vegamag and ABmag, for comparison with F12. PSF FWHM is the full width at half maximum of the point spread function, as derived in this work. All the indicated errors are at 1 $\sigma{}$.}
\end{flushleft}
\end{center}
\end{table*}
We re-analyzed the {\it HST} data presented by F12 (GO program 12256). The data were acquired with the Advanced Camera for Surveys (ACS) Solar Blind Camera (SBC) in the F140LP filter (far UV, FUV), and with the Wide Field Camera 3 (WFC3) in the F300X (near UV, NUV), F390W (C), F555W (V), F775W (I), and F160W (H) filters. 
We retrieved from the {\it HST} archive the final drizzled combined images, calibrated through the 6.1.0 version of the CALACS pipeline (16-May-2011) and through the 2.7  version of the  CALWF3 pipeline (21-May-2012).

Aperture photometry at the position of the HLX-1 counterpart was performed in the different bands using the {\it phot} task in the IRAF environment\footnote{IRAF is distributed by the National Optical Astronomy Observatories, which are operated by AURA, Inc., under cooperative agreement with the National Science Foundation.}.
The counts were measured within circular apertures of radius 0.4 arcsec in FUV and NUV. 
In the redder C, V, and I bands, because of the strong contamination from the S0 galaxy, we adopted a smaller aperture of 0.2 arcsec, and then applied aperture corrections from 0.2 to 0.4 arcsec computed from the most isolated 
and brightest stars in the frames. 
Finally, the instrumental magnitudes were calibrated into the {\it HST} Vegamag system applying the ACS and WFC3 photometric zeropoints available at {\it HST} webpages\footnote{{\tt http://www.stsci.edu/hst/acs/analysis/zeropoints/\#sbc} 
and {\tt http://www.stsci.edu/hst/wfc3/phot\_zp\_lbn}}. 
 In the H band, where the strong diffuse emission from the S0 galaxy requires a careful background subtraction, we first created a smoothed S0 `galaxy image' applying a Gaussian smoothing with $\sigma{}= 0.4$ arcsec to the original image, and then subtracted the smoothed image to the original frame. The value recovered with this procedure ($m=23.30 \pm 0.26$, in Vegamag) is consistent with the one obtained by F12 (i.e. $m=23.15 \pm 0.30$, in Vegamag).
 We provide in Table~1 both the magnitudes within a 0.4 arcsec aperture [$\approx$4 times the point spread function (PSF) full width at half maximum (FWHM)],
and the `total magnitudes' obtained by applying aperture corrections to infinity.  For the WFC3, these corrections  are available at the WFC3 photometric zeropoint page, while for ACS/SBC we computed it from the Tiny Tim PSF 
models\footnote{http://www.stsci.edu/hst/observatory/focus/TinyTim} (Krist 1995). Notice that the correction to infinity assumes that the source is point-like. 
Thus, we may be underestimating the total brightness of the source if extended tails, too faint to be detected against the strong galaxy background, are present. Once converted into the ABmag system, our magnitudes are in good agreement with those provided by F12.
\begin{table}
\begin{center}
\caption{Initial conditions of the $N-$body simulations: masses and scale lengths.}
 \leavevmode
\begin{tabular}[!h]{ccc}
\hline
Model galaxy properties & Primary & Secondary \\
\hline
DM Mass$^{\rm a}$ [$10^{11}$ M$_\odot{}$]          & 7.0, 12.5  & 0.3, 0.6\\
$M_\ast{}$$^{\rm b}$ [$10^{10}$ M$_\odot{}$]      & 7.0   & 0.2\\
$f_{\rm b/d}$                               &  0.25 & 0.25 \\
Gas Mass$^{\rm c}$ [$10^{8}$ M$_\odot{}$]    & 0 &  1.38 \\
Halo scale length$^{\rm d}$ [kpc] & 6.0 & 3.0\\
Disc scale length [kpc] & 3.7 & 3.0 \\
Disc scale height [kpc] & 0.37 & 0.30 \\
Bulge scale length [kpc] & 0.6 & 0.6 \\
\noalign{\vspace{0.1cm}}
\hline
\end{tabular}
\begin{flushleft}
\footnotesize{$^{\rm a}$  The mass of the DM halo in the primary (secondary) galaxy is $1.25\times{}10^{12}$ M$_\odot{}$ ($6\times{}10^{10}$ M$_\odot{}$) in run D and  $7\times{}10^{11}$ M$_\odot{}$ ($3\times{}10^{10}$ M$_\odot{}$) in all the other simulations.  $^{\rm b}$  $M_\ast{}$ is the total stellar mass of the galaxy (including both bulge and disc). $f_{\rm b/d}$  is the bulge-to-disc mass ratio. \\$^{\rm c}$The primary has no gas, while the gas of the secondary is distributed according to an exponential disc, with the same parameters (scale length and height) as the stellar disc. $^{\rm d}$ We name halo scale length the NFW scale radius $R_{\rm s}\equiv{}R_{200}/c$, where $R_{200}$ is the virial radius of the halo (NFW 1996) and $c$ the concentration (here we assume $c=12$ for both galaxies).}
\end{flushleft}
\end{center}
\end{table}

In the last column of Table~1, we also list the PSF FWHMs measured from the brightest and  most isolated stars in the images. Since bright point-like sources are absent in our FUV image, the FWHM in F140LP was obtained from SBC observations of the globular cluster NGC~6681 (program 9563).

The surface brightness profiles in five different filters, from F140LP to F775W, were derived performing photometry within circular apertures of increasing radii (up to 1 arcsec), and computing the background in an annulus of  width 0.08 arcsec at $r>1$ arcsec. To account for the uncertainties due to the highly variable S0 background, we repeated the computation varying the position of the background annulus between 1.0 and 1.4 arcsec in steps of 0.2 arcsec, and then averaging among the results (in the following, we refer to this procedure as first approach).  The estimated error on each circular aperture accounts for both the photometric uncertainty on the source and the fluctuations of the background. The flux in each annulus was obtained as the difference between the fluxes of two adjacent circular apertures. To derive the error in each annulus, the flux errors on the two adjacent circular apertures were summed in quadrature. The reported error bars are at 1 $\sigma{}$.

In F555W and F775W, because of the high S0 background, we attempted also a different approach, and tried to compute the profiles after subtraction of the galaxy emission. To this purpose, we first created a `galaxy image' by replacing the 
HLX-1 counterpart with background values computed in an adjacent annulus (task {\it imedit} in IRAF), and then by applying a Gaussian smoothing with $\sigma=0.4$ arcsec  to the image.  We then subtracted the galaxy image to the original frame, and computed the profiles following the same approach as before (in the following, we refer to this procedure as second approach).  The errors were computed in the same way as for the first approach. We notice that the background in this second approach is smoother by construction: the subtraction of a Gaussian smoothed image might smear out possible faint irregular features. This might introduce a systematic error, which is not accounted in the reported error bars. Thus, the errors on the second approach are likely underestimated. The differences between the two approaches will be further discussed in Sections~\ref{sec:surface} and ~\ref{sec:simul}.

We do not attempt to extract a profile from the F160W filter, as the contamination by the S0 is too high (the flux of the S0 galaxy is about 10 times higher than the flux of the HLX-1 counterpart). The reported magnitudes and surface brightness profiles for the counterpart of HLX-1 were not corrected for the reddening.
\begin{table*}
\begin{center}
\caption{Initial conditions of the $N-$body simulations: orbital properties.}
 \leavevmode
\begin{tabular}[!h]{ccccccccc}
\hline
  Run &  $b$  & $v_{\rm rel}$ & $\theta{}$, $\phi{}$, $\psi{}$ & $D$ & $E_{\rm s}$  & $L_{\rm s}$ & $e$  & Orbit spin \\
  &  (kpc) & (km s$^{-1}$) & (rad)  & (kpc) & (10$^4$ km$^2$ s$^{-2}$) & (10$^3$ km s$^{-1}$ kpc) & & \\
\hline
A        & 10.0   & 200 & $\pi{}/2$, $\pi{}$, 0 & 200 & 0.38  & 2.0 &  1.003 &  prograde \\
B        & 10.2   & 100 & $\pi{}/2$, 0, 2.94 & 150  & -1.65 & 1.0 &  0.997 &  retrograde\\
C        & 10.2   & 50  &  $\pi{}/2$, 0, 2.94 & 150 & -2.03 & 0.5 & 0.999 &  retrograde\\
D        & 10.2   & 100 &  $\pi{}/2$, 0, 2.94 & 150 & -3.25 & 1.0 & 0.998 &  retrograde\\
E1       & 30.1   & 100 & $-\pi{}/2$, 0, 2.94 & 150 & -1.66 & 3.0 & 0.971 & prograde \\
E2       & 30.1   & 100 & $\pi{}/2$, 0, 2.94  & 150 & -1.66 & 3.0 & 0.971  & retrograde\\
\noalign{\vspace{0.1cm}}
\hline
\end{tabular}
\begin{flushleft}
\footnotesize{$b$ and  $v_{\rm rel}$ are the impact parameter and the relative velocity (between the CMs of the two galaxies) at the initial distance, respectively. For the definition of $\theta{}$, $\phi{}$, $\psi{}$, see figure~1 of Hut \&{} Bahcall (1983). In particular, $\theta{}$ is the angle between the relative velocity vector ${\bf v}_{\rm rel}$ and the symmetry axis of the primary disc, $\phi{}$ describes the orientation of  ${\bf v}_{\rm rel}$ projected in the plane of the primary disc and $\psi{}$ describes the orientation of the initial distance vector ${\bf D}$ (between the CMs of the two galaxies) in the plane perpendicular to ${\bf v}_{\rm rel}$.\\
 $E_{\rm s}$ is the  specific orbital energy, i.e. the total energy divided by the reduced mass $\mu{}=m_1\,{}m_2/(m_1+m_2)$ (where $m_1$ and $m_2$ are the mass of the primary and of the secondary galaxy, respectively). $E_{\rm s}\equiv{}-G\,{}M/D+\,{}v_{\rm rel}^2/2$, where $M=m_1+m_2$ is the total mass of the two galaxies, 
$G$ is the gravitational constant and $D$ the initial distance between the CMs. \\$L_{\rm s}$ is the modulus of the specific orbital angular momentum, i.e. the angular momentum divided by the reduced mass. \\$e$ is the eccentricity ($e=[1+2\,{}E_{\rm s}\,{}L_{\rm s}^2/(G\,{}M)^2]^{1/2}$). \\An orbit is classified as prograde/retrograde depending on the alignment/counter-alignment of the orbital angular momentum of the secondary galaxy with respect to the spin of the primary galaxy.}
\end{flushleft}
\end{center}
\end{table*}

\subsection{$N-$body simulations}~\label{sec:nbody}
\begin{figure}
\center{{
\epsfig{figure=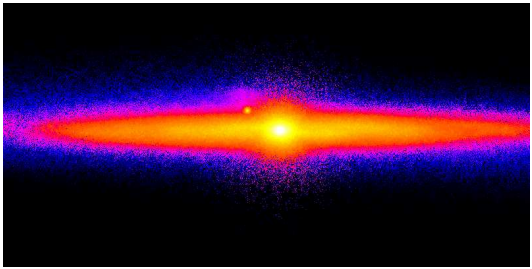,width=7cm} 
}}
\caption{\label{fig:fig1}
Projected mass density of stars in run~C at $t=2.0$ Gyr after the first pericentre passage. The primary galaxy is seen edge-on. The scale of the color-coded map is logarithmic, ranging from  $2.2\times{}10^{-6}$ M$_\odot{}$ pc$^{-3}$ to $22$ M$_\odot{}$ pc$^{-3}$.  The frame measures $50\times{}25$ kpc. 
}
\end{figure}
\begin{figure}
\center{{
\epsfig{figure=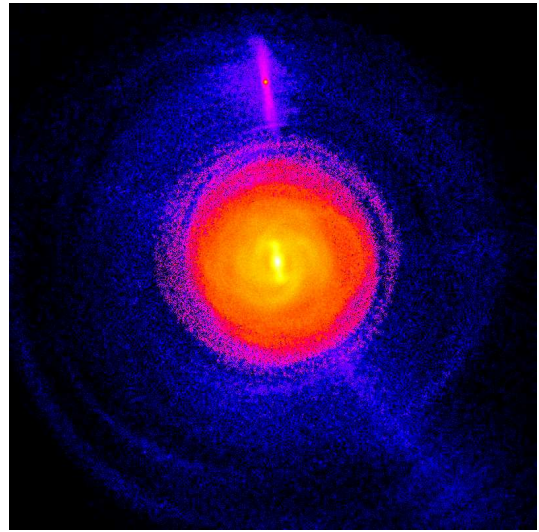,width=7.0cm} 
}}
\caption{\label{fig:fig2}
Projected mass density of stars in run~C at $t=2.0$ Gyr after the first pericentre passage.  The scale of the color-coded map is logarithmic, ranging from  $2.2\times{}10^{-6}$ M$_\odot{}$ pc$^{-3}$ to $22$ M$_\odot{}$ pc$^{-3}$. The primary galaxy is seen face-on. The frame measures $130$ kpc per edge. 
}
\end{figure}
In this paper, we compare the results of the photometric analysis with the outcomes of $N-$body simulations, modelling the merger between a S0 galaxy (which matches the properties of ESO 243-49) and a smaller (mass ratio 1:20) bulgy disc galaxy. In particular, we consider six simulations, labelled as run A, B, C, D, E1 and E2 (see Tables 2 and 3 for their properties). Simulations A and B were presented already in paper~I, while the remaining four runs are new.

As in paper~I, the initial conditions for both the primary galaxy and the secondary galaxy in the $N-$body model are generated by using an upgraded version of the code described in Widrow, Pym \&{} Dubinski (2008; see also Kuijken \&{} Dubinski 1995 and Widrow \&{} Dubinski 2005). The code generates self-consistent disc-bulge-halo galaxy models, derived from explicit distribution functions for each component, that are very close to equilibrium. In particular, the halo is modelled with a Navarro, Frenk \&{} White (1996, NFW) profile. We use an exponential disc model (Hernquist 1993), while the bulge is spherical and comes from a generalization of the Sersic law (Prugniel \&{} Simien 1997; Widrow et al. 2008).
 
Both the primary and the secondary galaxy have a stellar bulge and a stellar disc.
 The giant S0 galaxy has no gas, whereas the secondary galaxy has an initial gas mass of $1.38\times{}10^{8}$ M$_\odot{}$, distributed according to an exponential disc. Therefore, the initial configuration of the secondary galaxy is consistent with a low-mass gas-rich disc galaxy. The total mass of the secondary is $\sim{}1/20$ of the mass of the primary, classifying the outcome of the interaction as a minor merger. The masses of the various components and the scale lengths of the simulated galaxies are listed in Table~2. The total stellar mass in the primary galaxy was chosen to match the best-fitting value obtained by S10 from the comparison between the observed SED of ESO~243-49 and the stellar population models, assuming metallicity $Z=0.02$ and a Kroupa (1998) initial mass function (IMF).
The baryonic content of the simulated galaxies is the same in all the presented runs. Run D has more massive dark matter (DM) halos with respect to the other five runs.

Table~3 shows the orbital properties (impact parameter, relative velocity, orientation angles and total energy) of the six runs. The main common orbital feature of the six runs is that the centre of mass (CM) of the secondary galaxy is 
assumed to lie approximately on the same plane  as the disc of the primary galaxy. This was required as the offset between the HLX-1 counterpart and the disc of the S0 galaxy is relatively small ($\sim{}0.8$ kpc).
The adopted orbits are nearly parabolic (Table~3), in agreement with predictions from cosmological simulations (Khochfar \&{} Burkert 2006).

In all the simulations apart from run~C, the particle mass in the primary galaxy is $2.5\times{}10^5$ M$_\odot{}$ and $5\times{}10^4$ M$_\odot{}$ for DM and stars, respectively. The particle mass in the secondary galaxy is  $2.5\times{}10^4$ M$_\odot{}$ for DM and $5\times{}10^3$ M$_\odot{}$ for both stars and gas. 
The softening length is $\epsilon{}=0.1$ kpc. Run~C has a factor of 5 higher mass resolution and softening length $\epsilon{}=0.03$ kpc.

As in paper~I, we simulate the evolution of the models with the $N-$body/smoothed particle hydrodynamics (SPH) tree code gasoline (Wadsley, Stadel \&{} Quinn 2004). Radiative cooling, SF
 and supernova (SN) blastwave feedback are enabled, as described in Stinson et al. (2006, 2009, see also Katz 1992). The adopted parameters for SF and feedback  are the same as used in recent cosmological simulations capable of forming realistic galaxies in a wide range of masses (e.g., Governato et al. 2010; Guedes et al. 2011), and in recent simulations of galaxy-galaxy collisions (Mapelli \&{} Mayer 2012).

Figs.~\ref{fig:fig1} and  \ref{fig:fig2} show the projected density of stars in run~C  at $t=2.0$ Gyr after the first pericentre passage. In particular, the simulated galaxies in Fig.~\ref{fig:fig1} are oriented so that the S0 galaxy matches our almost edge-on view of ESO~243-49 and the projected position of the satellite remnant is very similar to the observed position of the HLX-1 counterpart (i.e. at $\sim{}3$ kpc from the S0 bulge and $\sim{}1$ kpc above the S0 disc).

In Fig.~\ref{fig:fig2}, the S0 galaxy is rotated face-on, so that the shells resulting from the ongoing merger are visible. In Fig.~\ref{fig:fig2}, the tidal features surrounding the nucleus of the stripped satellite are also evident.  The formation of the bar visible in the S0 galaxy was triggered by the interaction, as our galaxy model is marginally stable against bar formation when run in isolation  (i.e. it does not develop a bar, unless it is perturbed). In Fig.~\ref{fig:fig2}, the CM of the satellite is 44 kpc far from the CM of the S0 galaxy.  As we stressed in paper~I, we cannot exclude on the basis of the observational constraints (in particular, of  the line-of-sight velocity estimated by Wiersema et al. 2010) that  the  HLX-1 counterpart lies  relatively far from the S0 and appears close to it because of projection effects.

 Soria et al. (2013) recently pointed out that the projected velocity offset between ES0~243-49 and the counterpart of HLX-1 may be higher ($\sim{}420$ km s$^{-1}$) than previously estimated by Wiersema et al. (2010, $\sim{}170$ km s$^{-1}$).  If confirmed, a velocity offset $\sim{}420$ km s$^{-1}$, very close to the escape velocity for the estimated mass of ES0~243-49, is quite difficult to justify for a young SC, while it is consistent with an ongoing merger.

This higher velocity estimate is still consistent with the six simulations presented in this paper. The maximum three-dimensional velocity offset  between the CMs of the two galaxies occurs when the satellite galaxy is at the pericentre of the orbit. Such maximum velocity offset  is $v_{\rm max}=630$, 550, 500, 710, 540 and 540  km s$^{-1}$,  for run A, B, C, D, E1 and E2, respectively\footnote{We note that $v_{\rm max}$ is significantly higher in run~D than in other runs with the same $v_{\rm rel}$, because the mass of the two galaxies is higher in such simulation. We also stress that the orbits decay slowly because of dynamical friction, but this effect is almost negligible during the simulations, as the orbits are nearly unbound.}. Since the adopted orbits have  $e\sim{}1$ (Table~3), the one-dimensional projected velocity ranges from almost zero to a value very close to the three-dimensional velocity, depending on the adopted line-of-sight. Thus, there is a significant portion of the parameter space for which the one-dimensional projected velocity offset between the two simulated galaxies is close to $420$ km s$^{-1}$, and, at the same time, the projected position matches our view of ESO~243-49 and of the HLX-1 counterpart. This is particularly true when the satellite is close to a pericentre passage ($\sim{}5-10$ kpc distance from the CM of the S0 galaxy).

\begin{figure*}
\center{{
\epsfig{figure=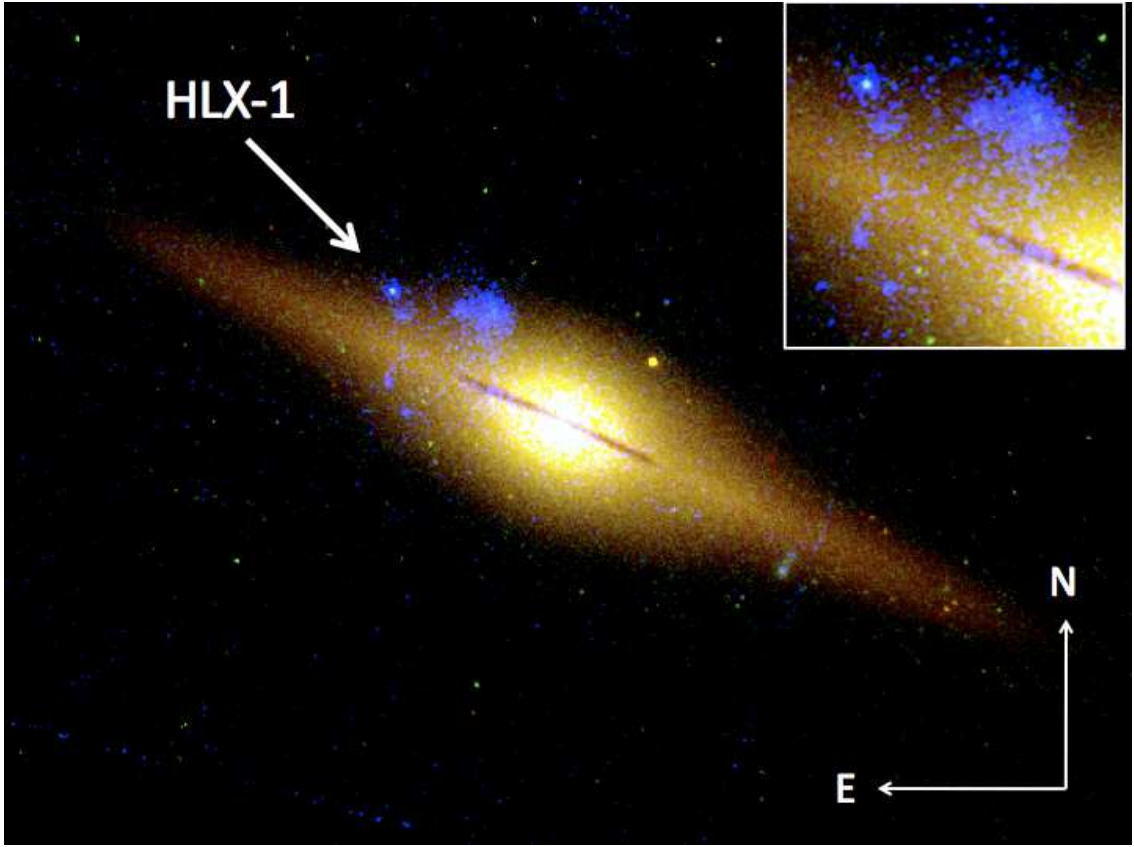,width=15.cm}  
\caption{\label{fig:fig3}
{Color-composite image of ESO~243-49. Red is F775W ($\lambda{}_{\rm pivot}=7733.6$ \AA{}), green is F390W ($\lambda{}_{\rm pivot}=3904.6$ \AA{}), blue is F140LP ($\lambda{}_{\rm pivot}=1527$ \AA{}). The arrow indicates the HLX-1 counterpart. North is up and east is left. The insert  is a zoom ($8\times{}8$ arcsec) around the HLX-1 counterpart and the background galaxy (BGG) at $z=0.03$. A Gaussian smoothing with $\sigma \sim 0.1$ arcsec (46.7 pc) has been applied to the FUV image.}}}} 
\end{figure*}

\section{The counterpart of HLX-1}~\label{sec:hlx1}
\subsection{An extended FUV emission?}~\label{sec:fuvextended}
The FUV, C, I color-combined image of ESO~243-49 is shown in Fig.~\ref{fig:fig3}. It shows an extended emission in FUV surrounding the HLX-1 counterpart. 
 Some structures, well visible in FUV, seem to connect the HLX-1 counterpart with the background galaxy (hereafter BGG) 
at $z\sim{}0.03$, which is at about half the projected distance between the bulge of ESO~243-49 and the HLX-1 counterpart  (Wiersema et al. 2010). This galaxy is likely part of a sheet of blue galaxies  at $z=0.03$, i.e. $\approx{}2000$ km s$^{-1}$ beyond the average recession velocity of Abell~2877, but superimposed to it because of projection effects (e.g. Malumuth et al. 1992). 

 In itself, this fact may lead to the conclusion that there is some connection between the counterpart of HLX-1 and the BGG. However, such a direct connection seems unlikely because the only line visible in the optical spectrum of the counterpart (interpreted as H$\alpha{}$) indicates a redshift for the HLX-1 counterpart closer to that of ES0~243-49 ($z=0.0224$) than to that of the BGG ($z\sim{}0.03$, Wiersema et al. 2010, see also Soria \&{} Hau 2012; Soria et al.  2013). While it is true that such spectral measurement relies on a rather delicate procedure of galaxy background subtraction, it is highly significant and, until new evidences coming from other spectral lines are found, we consider it reliable.

\begin{figure}
\center{{
\epsfig{figure=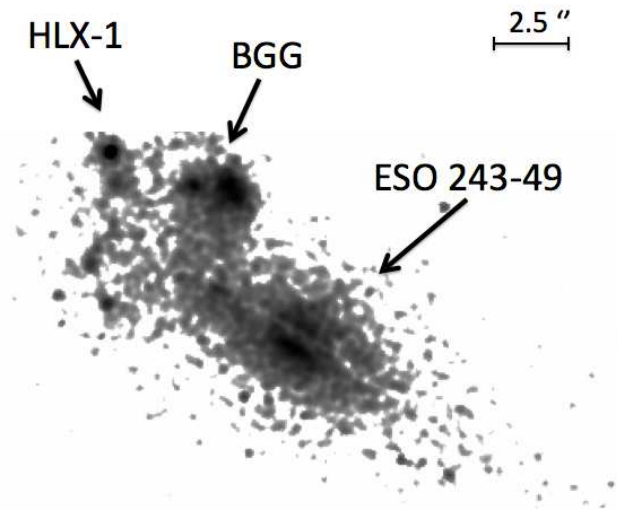,width=9.cm}  
\caption{\label{fig:fig4}
{FUV (F140LP) image of ESO~243-49.  North is up and east is left. A Gaussian smoothing with $\sigma \sim 0.2$ arcsec (93.3 pc) has been applied to the image. The arrows indicate (from left to right) the HLX-1 counterpart, the BGG and the bulge of ESO 243-49.}}}} 
\end{figure}

The extended structures connected with the HLX-1 counterpart are even more apparent in the smoothed FUV image (Fig.~\ref{fig:fig4}). Furthermore, this Figure shows that the BGG has two peaks in FUV and has a rather complicated morphology in the very young stellar component. From FUV photometry alone, it is hard to distinguish which extended structures belong to the HLX-1 counterpart and which belong to the BGG or even to the disc of ESO~243-49. However, it seems unlikely that none of these structures is physically connected with HLX-1.

\begin{figure}
\center{
\epsfig{figure=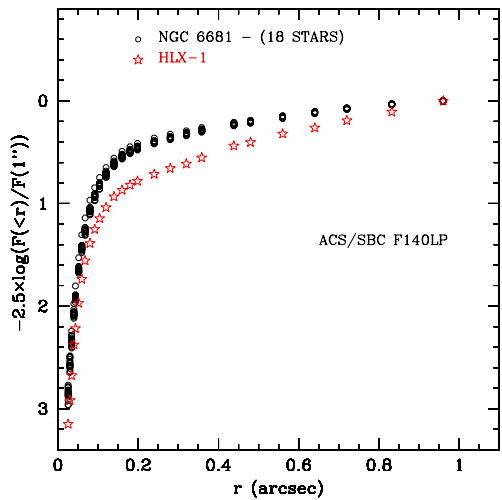,width=7.cm}  
\caption{\label{fig:fig5}
{Open circles: FUV flux profiles of 18 different stars in NGC~6681. These were used to measure the FUV PSF. Stars (red on the web): FUV flux profile of the counterpart of HLX-1.}}} 
\end{figure}

 Fig.~\ref{fig:fig5} compares the profiles of 18 stars in NGC~6681 (used to estimate the PSF, see Section~\ref{sec:data}) with the profile of the counterpart of HLX-1, both observed with the ACS/SBC F140LP filter. The profile of the counterpart of HLX-1 deviates significantly from the profile of a star. 
 On the other hand, we did the PSF fitting directly on the FUV image of the HLX-1 counterpart, and we found that the PSF FWHM estimated from the HLX-1 counterpart is consistent with 0.11 arcsec (i.e. with the  PSF FWHM obtained from the stars in NGC~6681, Table~1). This result, combined with Fig.~\ref{fig:fig5}, indicates that the profile of the HLX-1 counterpart is consistent with a point-like source (at the distance of ESO~243-49), surrounded by some faint extended emission. Again, such faint extended emission might either come from tidal tails that are physically associated with the HLX-1 counterpart, or be the effect of a mere superposition between the HLX-1 counterpart and the extended FUV emission of the BGG and/or of ESO~243-49.

\subsection{The observed surface brightness profiles}~\label{sec:surface}
\begin{figure*}
\center{{
\epsfig{figure=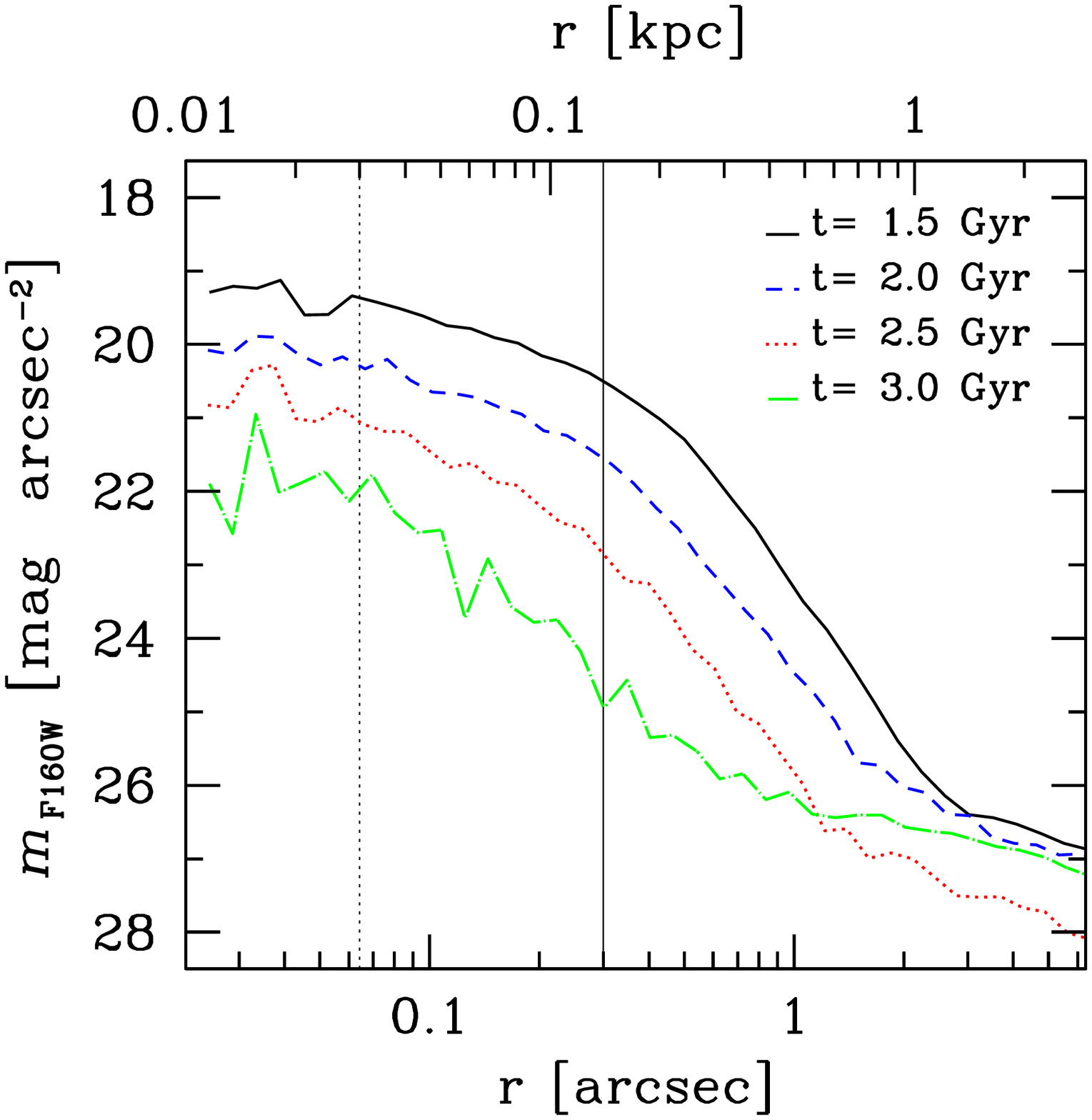,width=7.0cm} 
\epsfig{figure=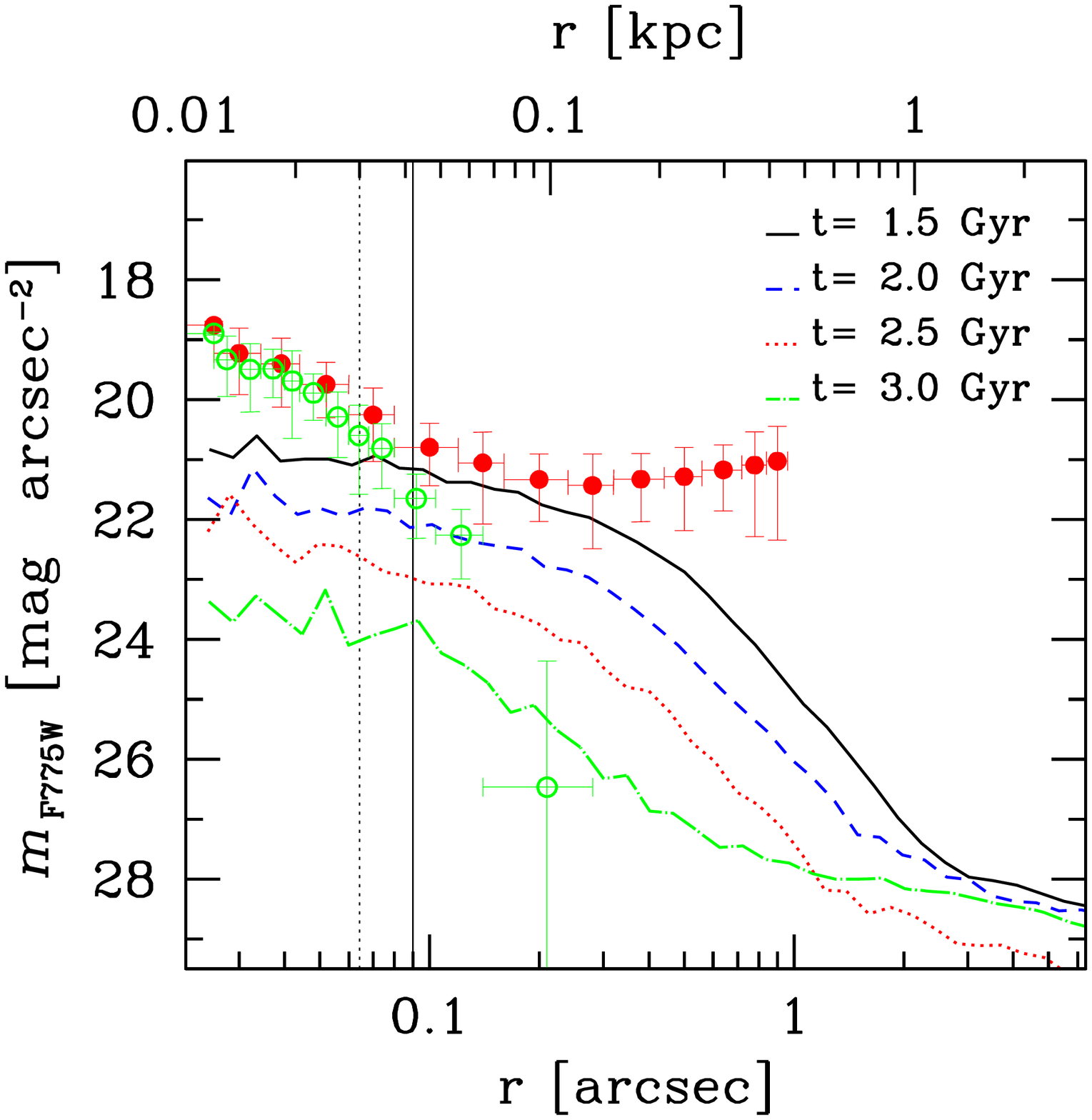,width=7.0cm} 
\epsfig{figure=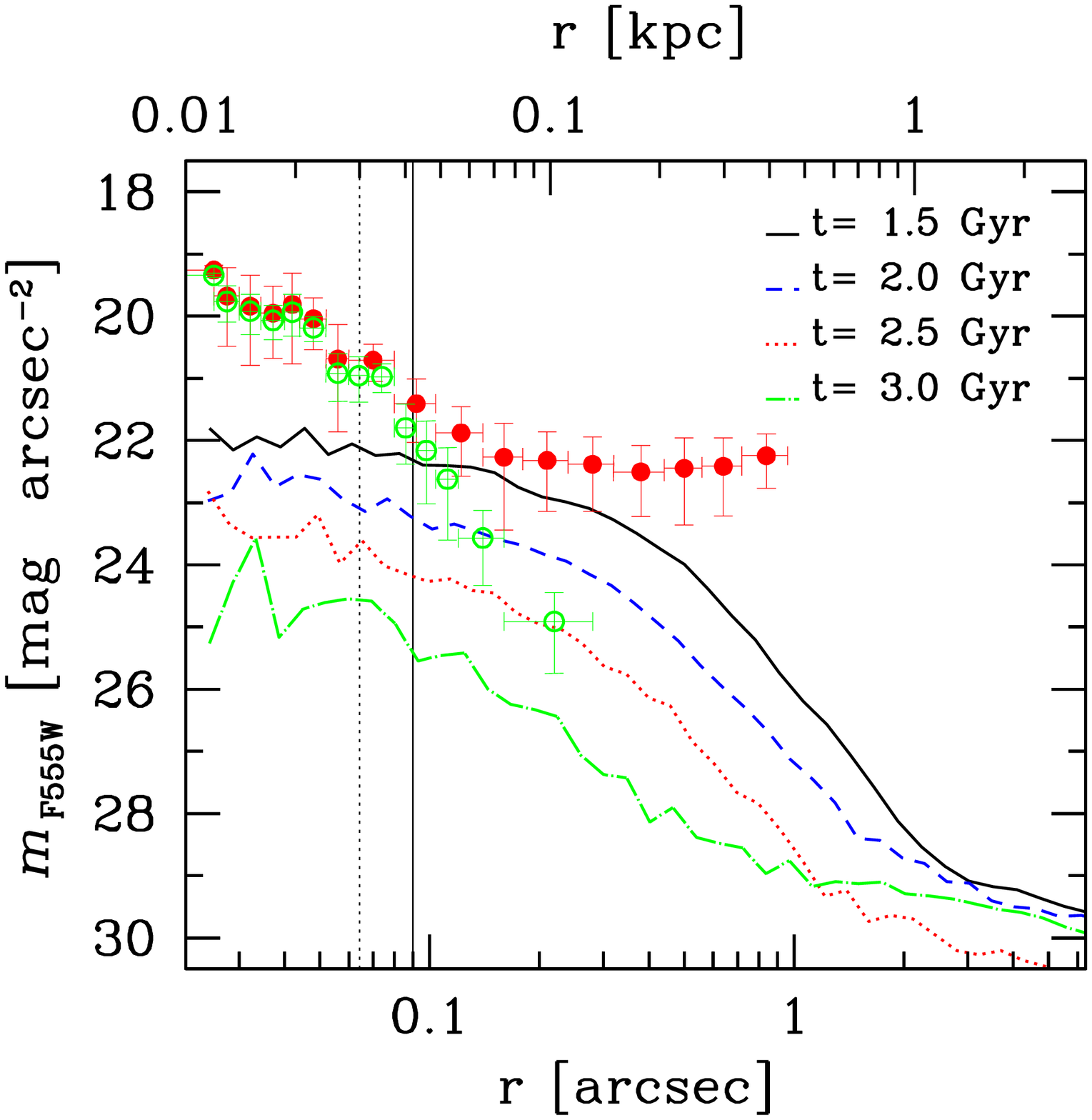,width=7.0cm} 
\epsfig{figure=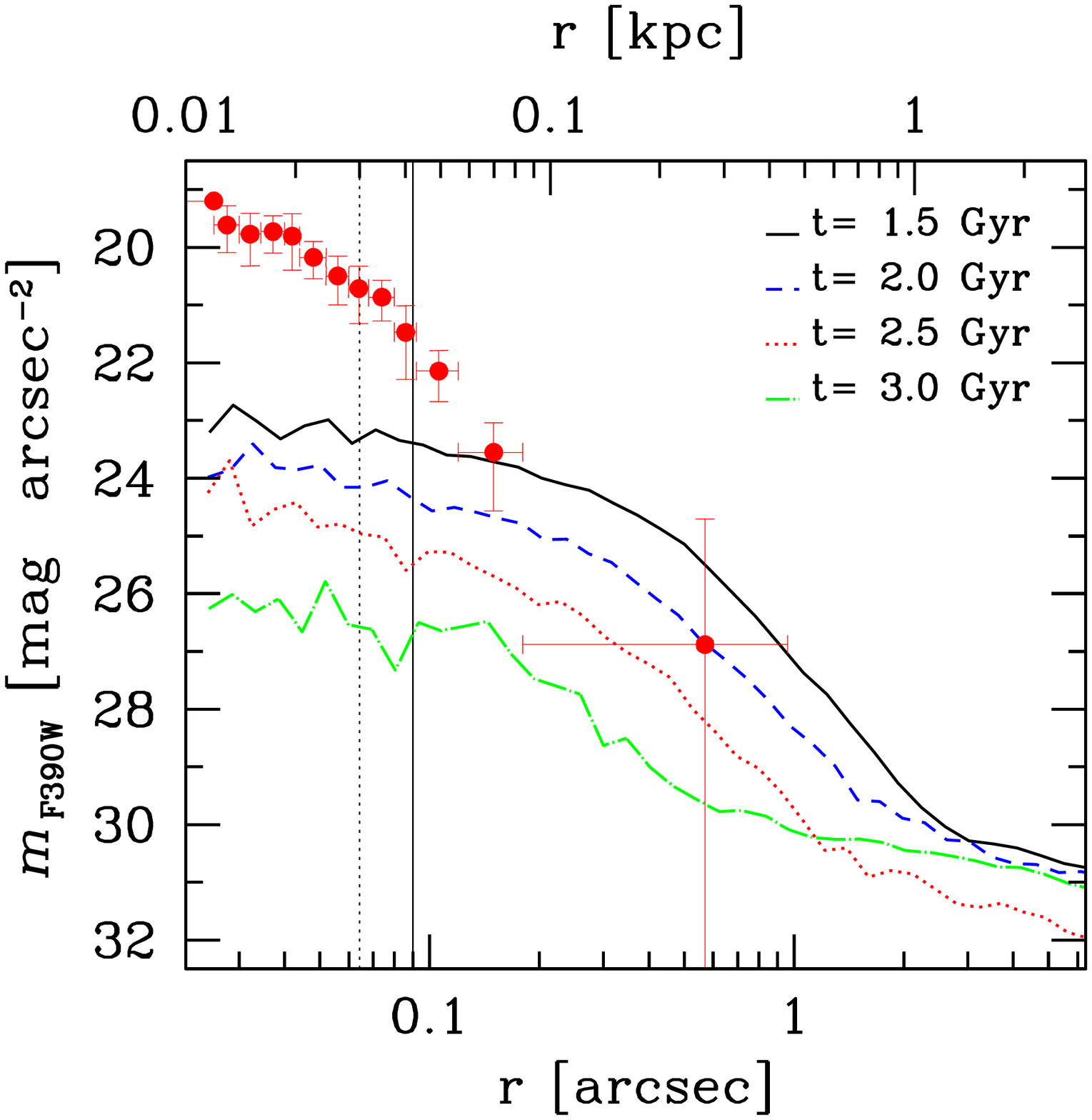,width=7.0cm} 
\epsfig{figure=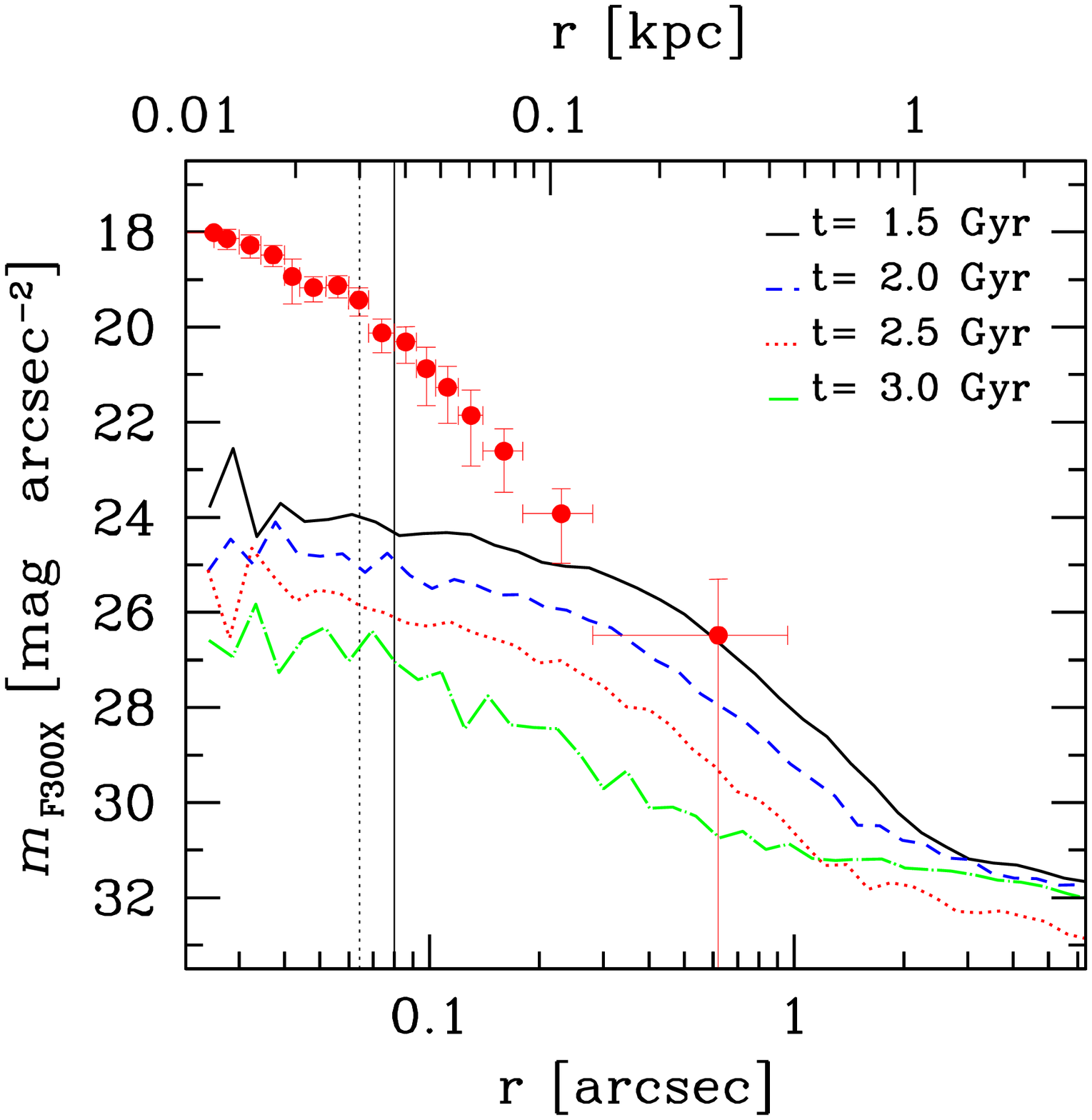,width=7.0cm} 
\epsfig{figure=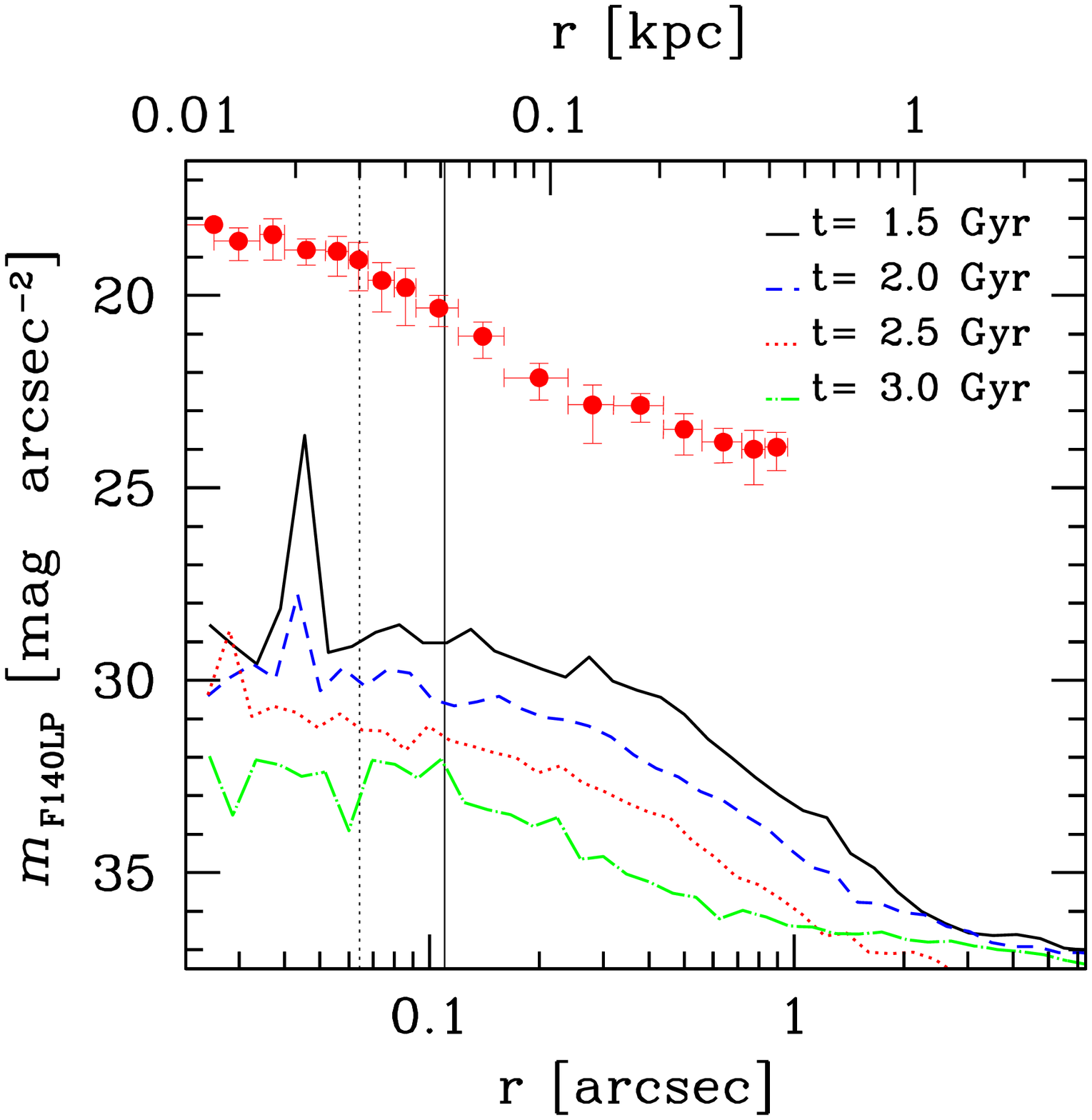,width=7.0cm} 
}}
\caption{\label{fig:fig6}
Surface brightness profiles of the HLX-1 counterpart in run~C (lines) and in the observations (circles). From left to right and from top to bottom: filter F160W, F775W, F555W, F390W, F300X and F140LP. In all panels, solid black line: simulated profile of the satellite galaxy at $t=1.5$ Gyr after the first pericentre passage; dashed line (blue on the web):  simulated profile at $t=2.0$ Gyr; dotted line (red on the web):  simulated profile at $t=2.5$ Gyr; dot-dashed line (green on the web):  simulated profile at $t=3.0$ Gyr. We assume E(B-V)=0 for the simulations.
Vertical dotted line: softening length. Vertical solid line: PSF FWHM (see Table~1).
We do not show the observed profile in F160W as the background of the S0 galaxy dominates over the flux of the HLX-1 counterpart in this filter. In the F775W and F555W filters, the filled red (open green) circles were obtained by subtracting the background according to the first (second) approach (as described in Section~\ref{sec:data}). 
In the other filters only the first approach was adopted. The errors are at 1 $\sigma{}$.
}
\end{figure*}

The circles in Fig.~\ref{fig:fig6} show the surface brightness profiles of the HLX-1 counterpart in five different filters, from F775W to F140LP (see Section~\ref{sec:data} for details on data analysis). We do not show the measurement in the F160W filter, as the S0 emission in this band is about 10 times higher than the flux of the HLX-1 counterpart, and we should rely on a model of the S0 flux to produce such profile. The S0 background is quite high even in F775W and in F555W, as it can be seen from the difference between filled and open circles in the corresponding two panels of  Fig.~\ref{fig:fig6}. 
As described in Section~\ref{sec:data}, the filled circles trace the profiles directly measured on the original frames (first approach), while the open circles show the photometry performed on a background subtracted image (second approach). In the first approach, the flattening of the profiles beyond $\sim{}$0.2 arcsec indicates where the S0 emission starts to dominate, preventing a robust derivation of the HLX-1 counterpart intrinsic profile. On the other hand, the subtraction of a Gaussian smoothed image adopted in the second approach might smear out all the features of an extended faint stellar halo.

The vertical lines in  Fig.~\ref{fig:fig6} show the FWHM of the PSF for each filter (see Table~1).
 In the F775W, F555W, F390W and F300X filters, there is no clear evidence for the HLX-1 counterpart to be associated with an extended source. In the FUV (F140LP) filter, the source appears sensibly more extended than implied by the PSF. This is an effect of the extended structures visible in the FUV image  (see Figs.~\ref{fig:fig3}, ~\ref{fig:fig4}, and the discussion in Section~\ref{sec:fuvextended}). On the basis of the available data, we cannot distinguish which extended FUV features are associated with HLX-1 or to the BGG or even to ESO~243-49. In case these features are associated with HLX-1, one of the possible explanations is that they are star forming tidal tails.

Given the lack of evidence for an extended emission, in Table~1 we model the counterpart as a point-like source. In particular, the magnitudes $m$ (in Vegamag) are the magnitudes extrapolated for an infinite aperture assuming that the source is point-like. These are in good agreement (within 0.2 mag) with the analysis by F12.

\subsection{Comparison with the simulations}~\label{sec:simul}
We can derive synthetic fluxes (in the six {\it HST} filters) from the $N-$body simulations, on the basis of each particle mass and age. In particular, we assume that the luminosity distance of ESO~243-49 is 96 Mpc (adopting the values of the cosmological parameters reported by Larson et al. 2011, i.e. Hubble parameter $H_0=71$ km s$^{-1}$ Mpc$^{-1}$, $\Omega{}_\Lambda{}=0.73$, $\Omega{}_{\rm M}=0.27$), and we use the single stellar population (SSP) models based on the tracks of Marigo et al. (2008), with the Girardi et al. (2010) case A correction for low-mass, low-metallicity asymptotic giant branch stars\footnote{\tt http://stev.oapd.inaf.it/cgi-bin/cmd\_2.3}. The tables of the SSP integrated magnitudes were implemented in the TIPSY visualization package for $N-$body simulations\footnote{\tt http://www-hpcc.astro.washington.edu/tools/tipsy/tipsy.html}. These SSP models are consistent with the ones adopted in F12 (see Maraston 2005 and Marigo et al. 2008 for details).

The lines in Fig.~\ref{fig:fig6} show the surface brightness profiles of the disrupted satellite in run~C (i.e. the highest resolution run), at different times and in different filters.  In Fig.~\ref{fig:fig6}, we assume  $A_{\rm V}=0$ for the simulated counterpart of HLX-1. Realistic values of $A_{\rm V}$ for the HLX-1 counterpart are $A_{\rm V}\sim{}0.04-0.18$ (S12). Thus, accounting for the reddening does not change our results significantly (see Section~\ref{sec:IDC}). 

The stars of the satellite that are present already in the initial conditions (i.e. that do not form during the interaction) 
were assumed to be 12 Gyr old. To generate the profiles in Fig.~\ref{fig:fig6}, we adopt a Kroupa IMF (Kroupa 1998) and metallicity $Z=0.008$ for the satellite galaxy. The main difference between our procedure ant that followed by F12 is that they assume a Salpeter IMF (Salpeter 1955), whereas we adopt a Kroupa IMF. This implies that the masses estimated by F12 are up to a factor of 3 higher than those estimated in our work. We discuss the effect of different IMFs in Section 3.3.1.

In  Fig.~\ref{fig:fig6}, we show the profiles only for run~C, as it has a higher spatial and mass resolution than the other five runs. This allows to reduce the fluctuations in the inner ($<0.1$ kpc) parts of the profile. On the other hand, the profiles obtained for the other runs have very similar shapes, although they are noisier in the central regions. We also re-simulated run~C with the same resolution as the other runs, to check the impact of resolution on the derived masses and magnitudes. We find differences smaller than $20$ per cent for $t<3$ Gyr since the last pericentre passage. At later times, the remnant of the satellite in the low-resolution run is disrupted faster than in the high-resolution run, because of the 
larger softening length. Thus, in the very late stages of the merger, low-resolution runs might underestimate the lifetime of the satellite remnant.

 The first thing we note from the comparison between observations and simulations in Fig.~\ref{fig:fig6} is that the simulations do not reproduce the observed profile at $r<0.1$ arcsec. This indicates that adding a further component (e.g. an irradiated disc component) is necessary to match the observations (see the discussion in Section~\ref{sec:IDC}). In this Section, we focus on the contribution of the simulated stellar population to the emission extending beyond $r\sim{}0.1$ arcsec.

The simulated profiles at $t=1.5-2$ Gyr after the first pericentre passage extend smoothly up to $\approx{}1$ kpc, before being tidally truncated by the interaction. At later times, the profiles are fast `eroded' by the interaction: $3$ Gyr after the first pericentre passage, the profile drops already at $\approx{}0.1$ kpc.

From the comparison with the observational profiles, we note that at sufficiently late times ($t>2.5$ Gyr after the first pericentre passage) the simulated profiles lie below the observed ones in all the filters considered.
Furthermore, we expect that it would be hard to disentangle the faint stellar halo of a disrupted satellite from the background of the S0 galaxy in the redder bands, even at earlier times.

The contribution of the stars formed after the beginning of the simulation is much smaller than that of the old stellar bulk. In particular, the simulated profiles  in the FUV filter are orders of magnitude fainter than the observed profile, confirming that UV filters are  dominated by disc irradiation.

 The spike in the simulated FUV profile at $t=1.5$ Gyr is due to a recent episode of SF. Given the sensitivity of the F140LP filter to recent SF and given the finite mass resolution of our simulations (the minimum mass of a newly formed star particle in run C is $5\times{}10^2$ M$_\odot{}$), even the formation of a single stellar particle in the inner $\approx{}100$ pc  produces significant fluctuations in the FUV profile.

In all the simulated profiles, the gravitational softening smears out any features within the softening length ($\epsilon{}=30$ pc for run C). Thus, the simulated profile within 30 pc has no physical meaning. Despite this, the  total mass enclosed within 30 pc can be compared with the data, as the observed PSF FWHM is always larger than (or similar to) the softening length, ensuring that the simulations and the observations are smoothed on the same scale.

In summary, the most important result from the simulated profiles is that the tidal tails are too faint to be distinguished in the {\it HST} images, provided that the merger is in a late stage ($>2.5$ Gyr from the first pericentre passage). Instead, it is highly unlikely that ESO 243-49 is in an earlier merger stage, unless the satellite galaxy is bulgeless and relatively low-surface density. In fact, pure disc satellites are disrupted faster than discs with bulges, due to their lower central density, which results in a smaller tidal radius (Gnedin, Hernquist \&{} Ostriker 1999; Feldmann, Mayer \&{} Carollo 2008).

\begin{table*}
\begin{center}
\caption{Evolution of the stellar mass bound to the satellite and corresponding predicted integrated magnitudes in the simulations (assuming a Kroupa IMF, $Z=0.008$ and E(B-V)=0).}
 \leavevmode
\begin{tabular}[!h]{lllllllllll}
\hline
 Run
 & {\it t} 
 & $N_{\rm p}$
 & $M_{<0.4''}$
 & F160W
 & F775W
 & F555W
 & F390W
 & F300X
 & F140LP \\
 & (Gyr)
 & 
 & ($10^6$ M$_\odot{}$)
 & (Vegamag)
 &  (Vegamag)
 &  (Vegamag)
 & (Vegamag)
 & (Vegamag) 
 & (Vegamag) \\
\hline
\noalign{\vspace{0.1cm}} 
A      & 1.0 & 1 & 64    & 19.5   & 21.1  & 22.1  & 23.0        & {\bf 23.0}  & {\bf 22.7} \\
       & 2.0 & 1 & 62    & 19.5   & 21.1  & 22.2  & 23.2        & {\bf 23.6}  & {\bf 23.5} \\
       & 3.0 & 2 & 40    & 20.0   & 21.6  & 22.7  & 23.9        & {\bf 24.6}  & {\bf 25.8} \\
       & 4.0 & 2 & 39    & 20.1   & 21.7  & 22.8  & 23.9        & {\bf 24.4}  & {\bf 24.5} \\
       & 5.0 & 3 & 26    & 20.5   & 22.1  & 23.2  & {\bf 24.4}  & {\bf  25.3} &  {\bf 29.4} \\
       & 6.0 & 4 & 15    & 21.2   & 22.7  & 23.9  & {\bf 25.0}  & {\bf  25.9} &  {\bf 31.3} \\
       & 7.0 & 4 & 15    & 21.2   & 22.7  & 23.9  & {\bf 25.0}  & {\bf  25.7} & {\bf 26.6} \\
\noalign{\vspace{0.2cm}} 
B      & 1.0 & 2 & 39    & 20.1         & 21.7        & 22.8        & 23.7       & {\bf 24.2} & {\bf 24.7} \\
       & 1.5 & 2 & 34    & 20.2         & 21.8        & 22.9        & {\bf 24.0} & {\bf 24.8} & {\bf 26.6} \\
       & 2.0 & 3 & 21    & 20.7         & 22.3        & 23.4        & {\bf 24.6} & {\bf 25.4} & {\bf 29.2} \\ 
       & 2.5 & 4 & 9.5   & 21.6         & 23.2        &  {\bf 24.3} & {\bf 25.5} & {\bf 26.3} & {\bf 31.4} \\ 
       & 3.0 & 4 & 8.6   & 21.7         & 23.3        &  {\bf 24.4} & {\bf 25.6} & {\bf 26.5} & {\bf 31.8} \\
       & 3.5 & 4 & 8.4   & 21.8         & 23.3        &  {\bf 24.5} & {\bf 25.6} & {\bf 26.5} & {\bf 31.9} \\  
       & 4.0 & 5 & 2.6   & 23.0         & {\bf 24.6}  &  {\bf 25.7} & {\bf 26.9} & {\bf 27.8} & {\bf 33.2} \\
       & 4.5 & 6 & 0.53  & {\bf 24.8}   & {\bf 26.4}  &  {\bf 27.5} & {\bf 28.6} & {\bf 29.6} & {\bf 34.9} \\
\noalign{\vspace{0.2cm}} 
C      & 1.0 & 2 & 28    & 20.4         & 22.0         & 23.1       & {\bf 24.2} & {\bf 25.0} & {\bf 27.7} \\
       & 1.5 & 3 & 15    & 21.1         & 22.7         & 23.8       & {\bf 24.9} & {\bf 25.8} & {\bf 29.8} \\
       & 2.0 & 4 & 6.1   & 22.1         &  {\bf 23.7}  & {\bf 24.8} & {\bf 25.9} & {\bf 26.8} & {\bf 31.8} \\
       & 2.5 & 5 & 2.2   &  {\bf 23.2}  &  {\bf 24.8}  & {\bf 25.9} & {\bf 27.1} & {\bf 28.0} & {\bf 33.2} \\
       & 3.0 & 6 & 0.57  &  {\bf 24.7}  &  {\bf 26.3}  & {\bf 27.4} & {\bf 28.5} & {\bf 29.4} & {\bf 34.8} \\
       & 3.5 & 7 & 0.21  &  {\bf 25.7}  &  {\bf 27.3}  & {\bf 28.4} & {\bf 29.6} & {\bf 30.5} & {\bf 35.9} \\
\noalign{\vspace{0.2cm}} 
D      & 1.0 & 2  & 33   & 20.2         & 21.8         & 22.9       & 23.9       & {\bf 24.5} & {\bf 25.7} \\
       & 1.5 & 4  & 12   & 21.2         & 22.8         & 23.9       & {\bf 25.0} & {\bf 25.8} & {\bf 27.8} \\
       & 2.0 & 5  & 6.8  & 21.9         &  {\bf 23.5}  & {\bf 24.6} & {\bf 25.7} & {\bf 26.6} & {\bf 30.8} \\
       & 2.5 & 6  & 3.2  & 22.7         &  {\bf 24.3}  & {\bf 25.4} & {\bf 26.6} & {\bf 27.4} & {\bf 32.6} \\
       & 3.0 & 7  & 0.63 & {\bf 24.6}   &  {\bf 26.1}  & {\bf 27.3} & {\bf 28.4} & {\bf 29.3} & {\bf 34.6} \\
\noalign{\vspace{0.2cm}}  
E1     & 1.0  & 1 & 62   & 19.6         & 21.2         & 22.3       & 23.4       & {\bf 24.2} & {\bf 25.8} \\
       & 2.0 & 2  &  43  & 20.0         & 21.6         & 22.7       & 23.8       & {\bf 24.6} & {\bf 26.5} \\
       & 3.0 & 4  &  18  & 20.9         & 22.5         & 23.6       & {\bf 24.7} & {\bf 25.5} & {\bf 26.7} \\
       & 4.0 & 5  & 9.7  & 21.6         & 23.2         & {\bf 24.3} & {\bf 25.5} & {\bf 26.4} & {\bf 31.6} \\
       & 5.0 & 6  & 4.0  & 22.6         &  {\bf 24.2}  & {\bf 25.3} & {\bf 26.5} & {\bf 27.4} & {\bf 32.7} \\
       & 5.5 & 7  & 1.5  & {\bf 23.7}   &  {\bf 25.3}  & {\bf 26.4} & {\bf 27.6} & {\bf 28.5} & {\bf 33.8} \\
       & 6.0 & 8  & 0.47 & {\bf 24.9}   &  {\bf 26.5}  & {\bf 27.6} & {\bf 28.8} & {\bf 29.7} & {\bf 35.1} \\
\noalign{\vspace{0.2cm}} 
E2     & 1.0  & 1  & 57 & 19.7          & 21.3         & 22.4       & 23.5       & {\bf 24.3} & {\bf 25.7} \\
       & 2.0 & 3 &  43  & 20.0          & 21.6         & 22.7       & 23.7       & {\bf 24.5} & {\bf 26.1} \\
       & 3.0 & 4  &  22  & 20.6         & 22.2         & 23.3       & {\bf 24.3} & {\bf 24.7} & {\bf 25.0} \\
       & 4.0 & 5  & 11  & 21.4          & 23.0         & {\bf 24.1} & {\bf 25.2} & {\bf 26.0} & {\bf 29.3} \\
       & 5.0 & 6 & 5.6  & 22.2          &  {\bf 23.7}  & {\bf 24.8} & {\bf 26.0} & {\bf 26.8} & {\bf 30.4} \\
       & 6.0 & 7 & 2.9  & 22.9          &  {\bf 24.5}  & {\bf 25.6} & {\bf 26.8} & {\bf 27.7} & {\bf 33.0} \\
       & 6.5 & 8 & 1.1  & {\bf 24.0}    &  {\bf 25.5}  & {\bf 26.6} & {\bf 27.8} & {\bf 28.7} & {\bf 34.0} \\
\noalign{\vspace{0.1cm}}
\hline
\end{tabular}
\begin{flushleft}
\footnotesize{{\it t}: time elapsed since the first pericentre passage;
$N_{\rm p}$: number of pericentre passages; $M_{<0.4''}$: stellar mass of the satellite within 0.4 arcsec. The values of F160W, F775W, F555W, F390W, F300X and F140LP indicate the integrated magnitudes (in Vegamag), in the central 0.4 arcsec of the simulated satellites, in the corresponding {\it HST} filters.  The bold face indicates simulated magnitudes consistent with (or higher than) the observed value of $m(<0.4'')$ (Table~1).
}
\end{flushleft}
\end{center}
\end{table*}

Table~4 reports the stellar mass within $r_{\rm cut}=0.4$ arcsec and the corresponding magnitudes (in the six different filters) for all the simulations, at the most relevant snapshots. $r_{\rm cut}$ was chosen to facilitate the comparison with the observed magnitudes, $m(<0.4'')$, reported in Table~1.
In Table~4, we also show the number of pericentre passages ($N_{\rm p}$) for each tabulated snapshot. $N_{\rm p}$ is found to be a crucial parameter, as most of the mass is stripped by the satellite remnant during each pericentre passage (see Table~4).
\begin{figure}
\center{{
\epsfig{figure=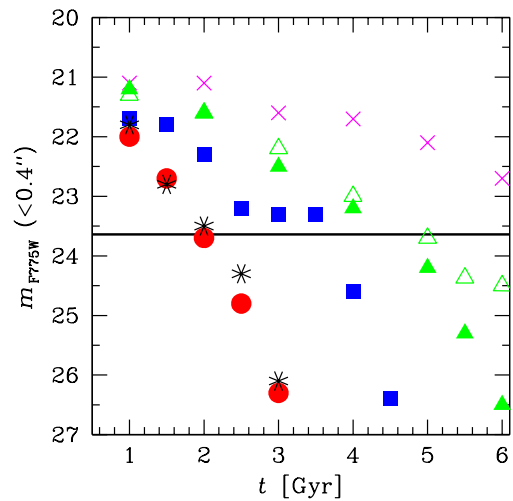,width=7cm} 
}}
\caption{\label{fig:fig7}
F775W magnitude within 0.4 arcsec in the simulated satellite galaxy, as a function of time. $t=0$ is the first pericentre passage. Crosses (magenta on the web): run~A; filled squares (blue on the web): run~B; filled circles (red on the web): run~C; black asterisks: run~D; filled triangles (green on the web): run~E1; open triangles (green on the web): run~E2. Horizontal solid line: observed  F775W magnitude within 0.4 arcsec (Table~1). We assume E(B-V)=0 for the simulations.}
\end{figure}
\begin{figure}
\center{{
\epsfig{figure=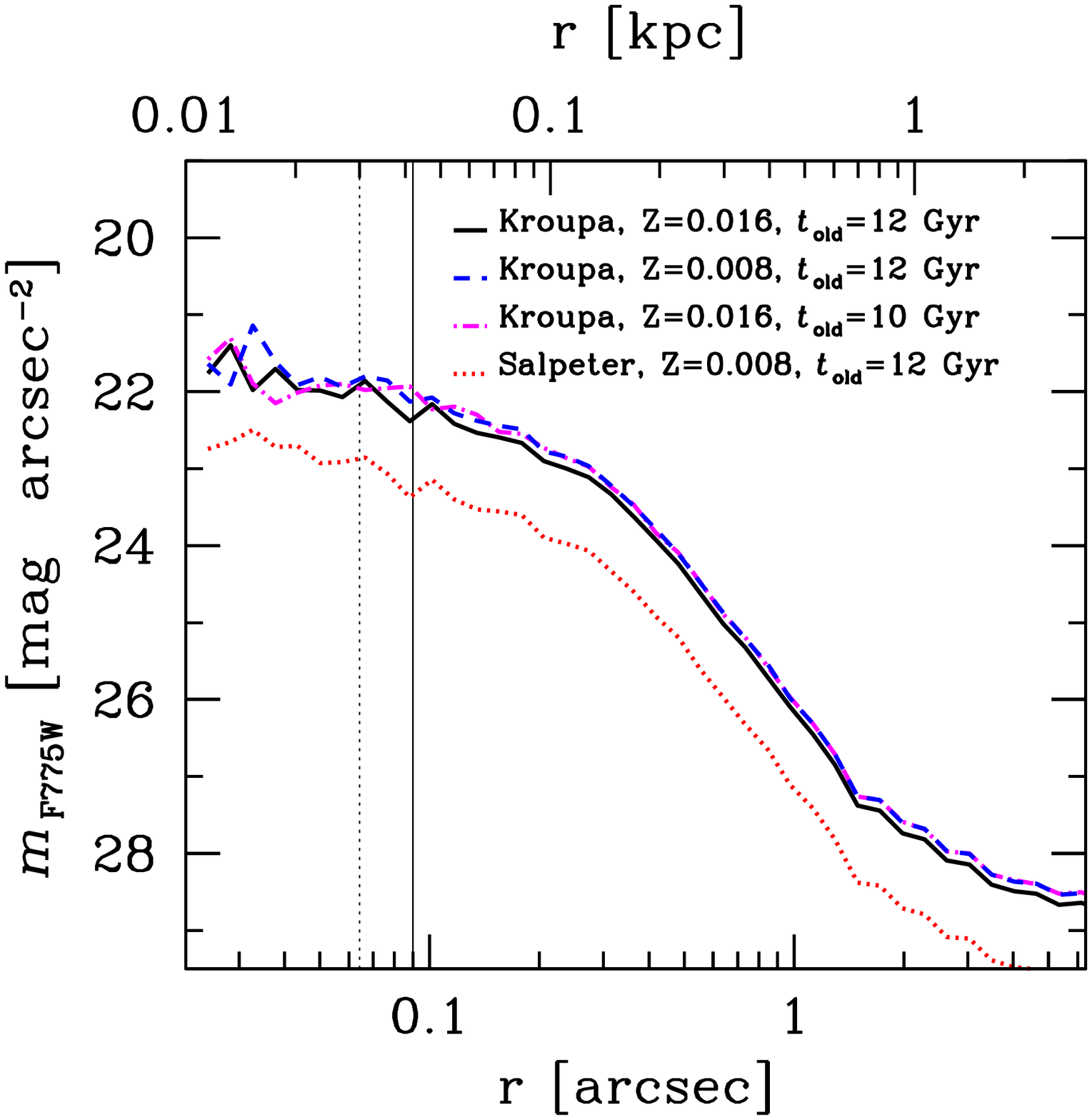,width=7cm} 
\epsfig{figure=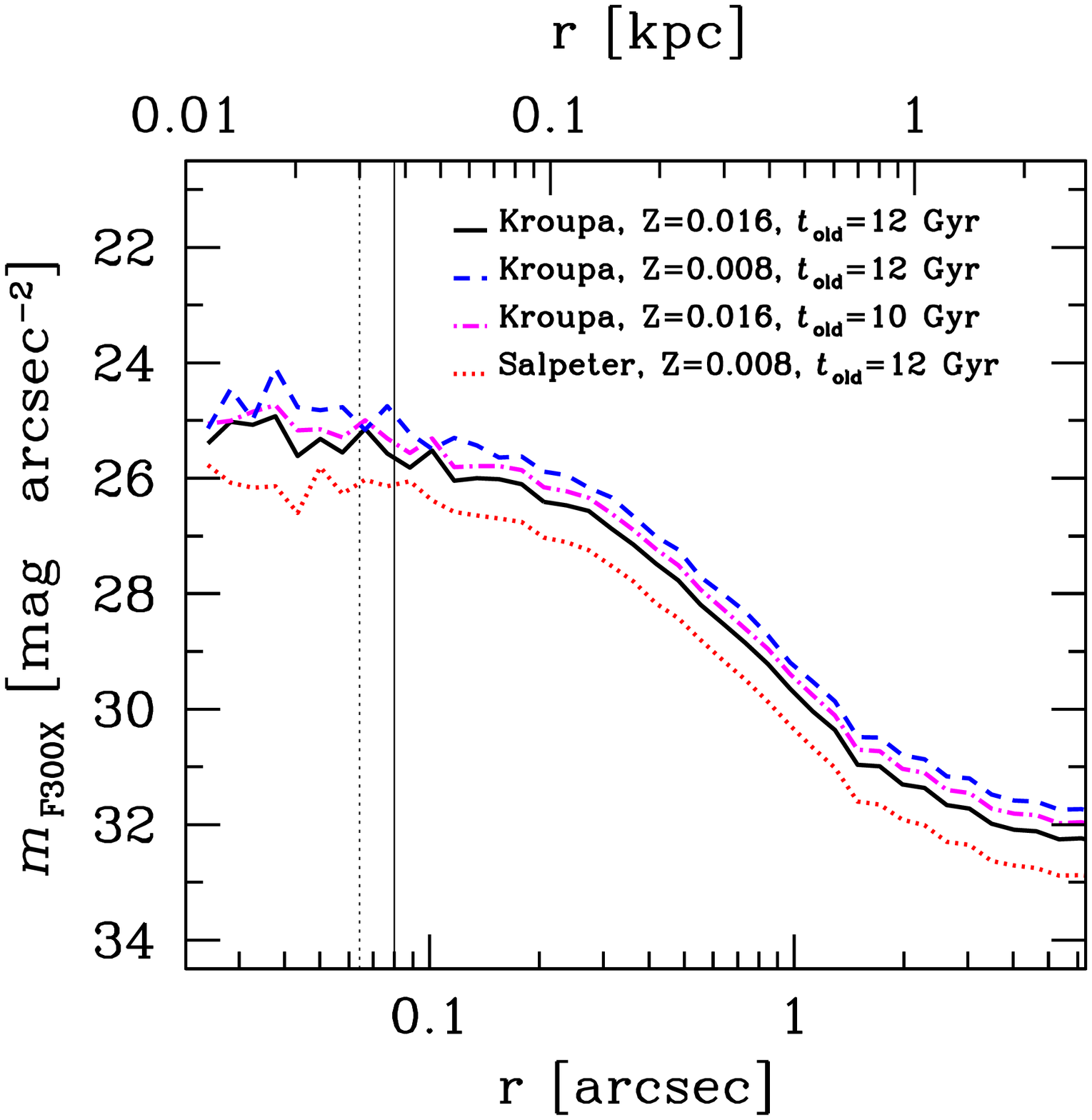,width=7cm} 
}}
\caption{\label{fig:fig8}
Surface brightness profile  of the satellite galaxy in the F775W filter (top) and in the F300X filter (bottom), for run~C, at $t=2.0$ Gyr since the first pericentre passage. Solid black line: Kroupa IMF, metallicity $Z=0.016$ and  age $t_{\rm old}=12$ Gyr for the bulk of the stellar population (i.e. formed before the merger); dashed line (blue on the web): Kroupa IMF, $Z=0.008$ and $t_{\rm old}=12$ Gyr; dot-dashed line (magenta on the web):  Kroupa IMF, $Z=0.016$ and $t_{\rm old}=10$ Gyr; dotted line (red on the web): Salpeter IMF,  $Z=0.008$ and $t_{\rm old}=12$ Gyr. We assume E(B-V)=0. The dashed line is the same as in Fig.~\ref{fig:fig6}.
}
\end{figure}

 The strongest constraints come from the integrated F160W magnitudes. In most runs, the  F160W magnitude is consistent with that of the HLX-1 counterpart [$m(<0.4'')=23.49$ in F160W, Table~1] only after $\gtrsim{}6$ pericentre passages. A magnitude $m(<0.4'')=23.49$ in the F160W filter corresponds to a stellar mass of the merger remnant $\sim{}1.7\times{}10^6$ M$_\odot{}$ (assuming a Kroupa IMF). Runs C and D (i.e. those with the most bound orbits, see Table~3) reach this value of $m(<0.4'')$ already $2.5-3$ Gyr after the first pericentre passage. Run~B satisfies this requirement only $\sim{}4$ Gyr after the first pericentre passage. Run~A does not match this requirement for the entire simulation (7 Gyr since the first pericentre passage). This indicates that the specific orbital energy ($E_{\rm s}$) of the satellite is a critical parameter to reproduce the properties of the HLX-1 counterpart.

Run~E2, which differs from run~B only for the larger impact parameter (see Table~3), satisfies the requirement that the F160W magnitude be higher than 23.49 only $6.0-6.5$ Gyr after the first pericentre passage, i.e. $\sim{}2$ Gyr later than run~B. Thus, the impact parameter of the satellite is another important ingredient of the merger scenario for HLX-1.

Run~E1 satisfies the requirement that the F160W magnitude be higher than 23.49 about half a Gyr earlier than Run~E2. The only difference between runs E1 and E2 is that, in the former run, the satellite has a prograde orbit, whereas in the latter the satellite has a retrograde orbit. In prograde orbits, the relative velocity between stars of the primary and stars of the secondary is smaller, increasing the tidal torques between the two discs and enhancing the disruption of the satellite (Toomre \&{} Toomre 1972). However, we stress that in our runs the differences between prograde and retrograde orbits are relatively minor, as the orbits are extremely radial and the passage through the primary disc is short if compared with the entire orbit of the secondary.

 Fig.~\ref{fig:fig7} shows the evolution with time of the simulated F775W magnitude for all the runs. The horizontal solid line is the observed value reported in Table~1. As already discussed for the results presented in Table~4,  Fig.~\ref{fig:fig7} confirms that run~C and run~D evolve in a similar way and are consistent with the observations of the HLX-1 counterpart already $\approx{}2$ Gyr after the first pericentre passage. Runs B, E1 and E2 evolve more slowly, whereas run~A is never consistent with the photometric constraints. We note that the observational constraints derived from the filter F775W are significantly less affected by the subtraction of the flux of the S0 galaxy than those derived from F160W (see the discussion in Sections~\ref{sec:data} and \ref{sec:surface}).

 The magnitudes provided in Table~4, as well as the simulated values shown in Figs.~\ref{fig:fig6} and \ref{fig:fig7} assume no absorption for the counterpart of HLX-1. From the simulations, we can estimate only the hydrogen column density due to the gas initially associated with the satellite galaxy, as the S0 galaxy was assumed to have initially no gas. Furthermore, in our simulations we do not have recipes for multi-phase gas. Then, we include in the estimate of the hydrogen column density also gas which may be molecular or ionized. Keeping these caveats in mind, the hydrogen column density for run C at $t=2.5$ Gyr is about $1-4\times{}10^{20}$ atoms cm$^{-2}$ (for the lines of sight that best match the observed position of HLX-1). Using the relation between optical extinction and hydrogen column density in  G\"uver \&{} \"Ozel (2009), we obtain $A_{\rm V}=0.045-0.18$, relatively small and consistent with the estimates by S12. The effects of correcting for this absorption will be discussed in Section~\ref{sec:IDC}.

\begin{table}
\begin{center}
\caption{Contribution to the magnitude inside 0.4 arcsec of the irradiated disc component (IDC), of the stellar population component (SPC) and of the sum of these two components (TOTAL), for the model of the HLX-1 counterpart.}
 \leavevmode
\begin{tabular}[!h]{llllll}
\hline
 Filter & IDC          & SPC          & TOTAL        & TOTAL & $m(<0.4'')$\\ 
        & $A_{\rm V}=0$  & $A_{\rm V}=0$ & $A_{\rm V}=0$ & $A_{\rm V}=0.18$ & \\
\hline 
 F140LP & 21.88        & 33.2         & 21.88        & 22.21 & $22.21 \pm 0.03$ \\
 F300X  & 22.55        & 28.0         & 22.54        & 22.86 & $22.80 \pm 0.05$\\
 F390W  & 23.76        & 27.1         & 23.71        & 23.97 & $24.04 \pm 0.05$\\
 F555W  & 24.13        & 25.9         & 23.94        & 24.13 & $24.11 \pm 0.05$\\
 F775W  & 24.04        & 24.8         & 23.60        & 23.72 & $23.64 \pm 0.15$\\
 F160W  & 24.12        & 23.2         & 22.82        & 22.86 & $23.49 \pm 0.26$\\
\noalign{\vspace{0.1cm}}
\hline
\end{tabular}
\begin{flushleft}
\footnotesize{All the reported magnitudes are in Vegamag. For the SPC we adopt the stellar population of the satellite galaxy in run~C at $t=2.5$ Gyr after the first pericentre passage. For the sum of the two components we show the values obtained assuming $A_{\rm V}=0$ and $A_{\rm V}=0.18$  in columns 4 and 5, respectively.  In the last column, we report the observed magnitude within 0.4 arcsec, $m(<0.4'')$, the same as shown in Table 1, to facilitate the comparison.}
\end{flushleft}
\end{center}
\end{table}

\subsubsection{Dependence on the assumptions}
To derive the simulated surface brightness profiles reported in Fig.~\ref{fig:fig6}, we made three fundamental assumptions: (i) Kroupa IMF; (ii) metallicity Z=0.008 (i.e. about half solar); (iii) age of the old stellar population $t_{\rm old}=12$ Gyr.

Fig.~\ref{fig:fig8} shows how strongly the results depend on these assumptions, considering an infrared (F775W) and an UV filter (F300X). In particular, in this figure we vary the three aforementioned assumptions, considering also a Salpeter (1955) IMF (with minimum mass 0.01 M$_\odot{}$), $Z=0.016$ (i.e. nearly solar metallicity) and $t_{\rm old}=10$ Gyr. The strongest differences depend on the choice of the IMF: a Salpeter IMF is about 1 mag fainter in both F775W and F300X filters. With a Salpeter IMF the maximum stellar mass consistent with the observed HLX-1 counterpart (assuming that $t_{\rm old}\ge{}10$ Gyr) is $6\times{}10^6$ M$_\odot{}$ (in agreement with F12).
Differences in the metallicity affect especially the UV filters but are within 0.5 magnitudes (unless the metallicity of the satellite galaxy is much lower than 0.5 Z$_\odot$).
\subsection{Adding disc irradiation}~\label{sec:IDC}
\begin{figure*}
\center{{
\epsfig{figure=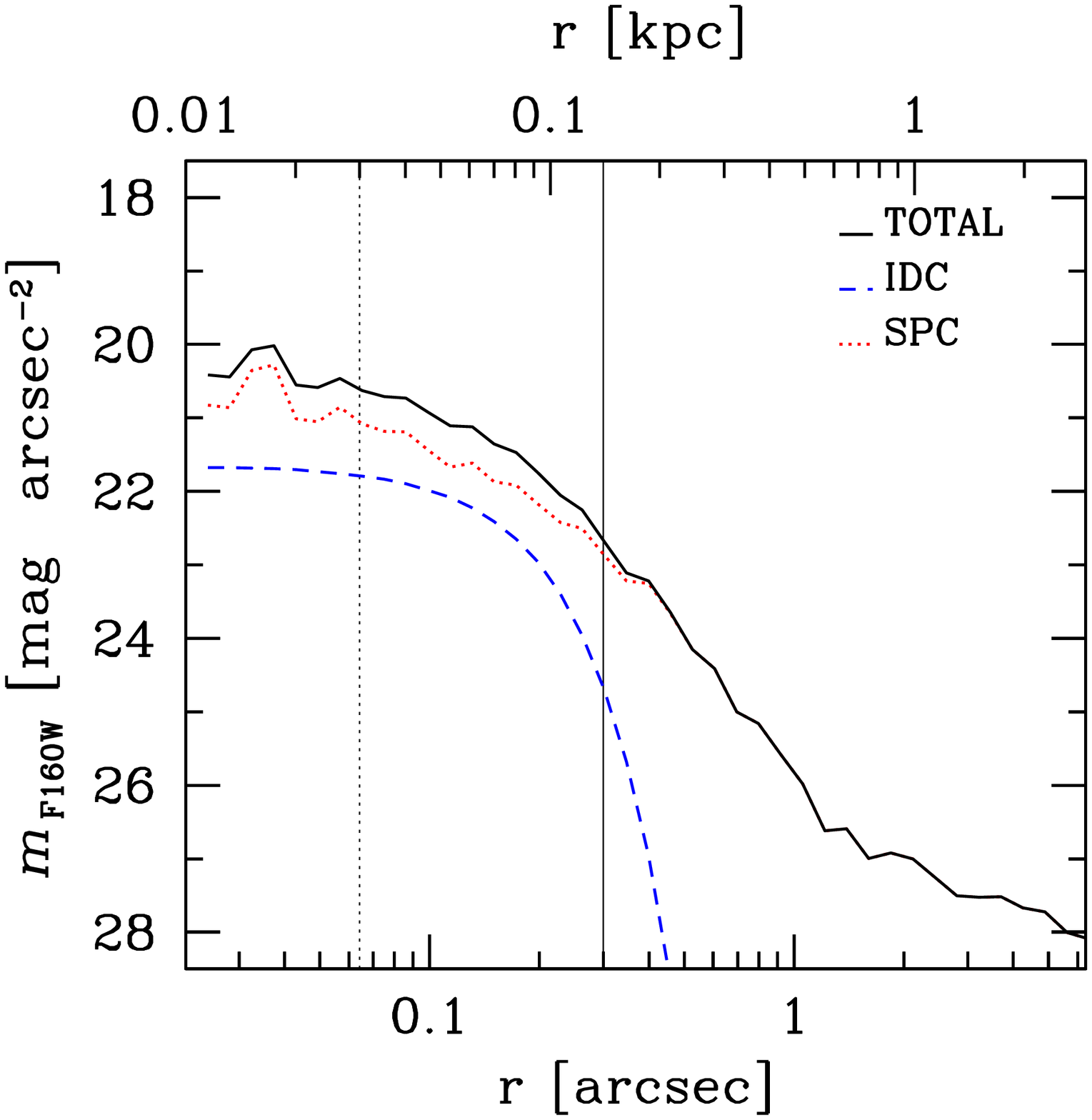,width=7.0cm} 
\epsfig{figure=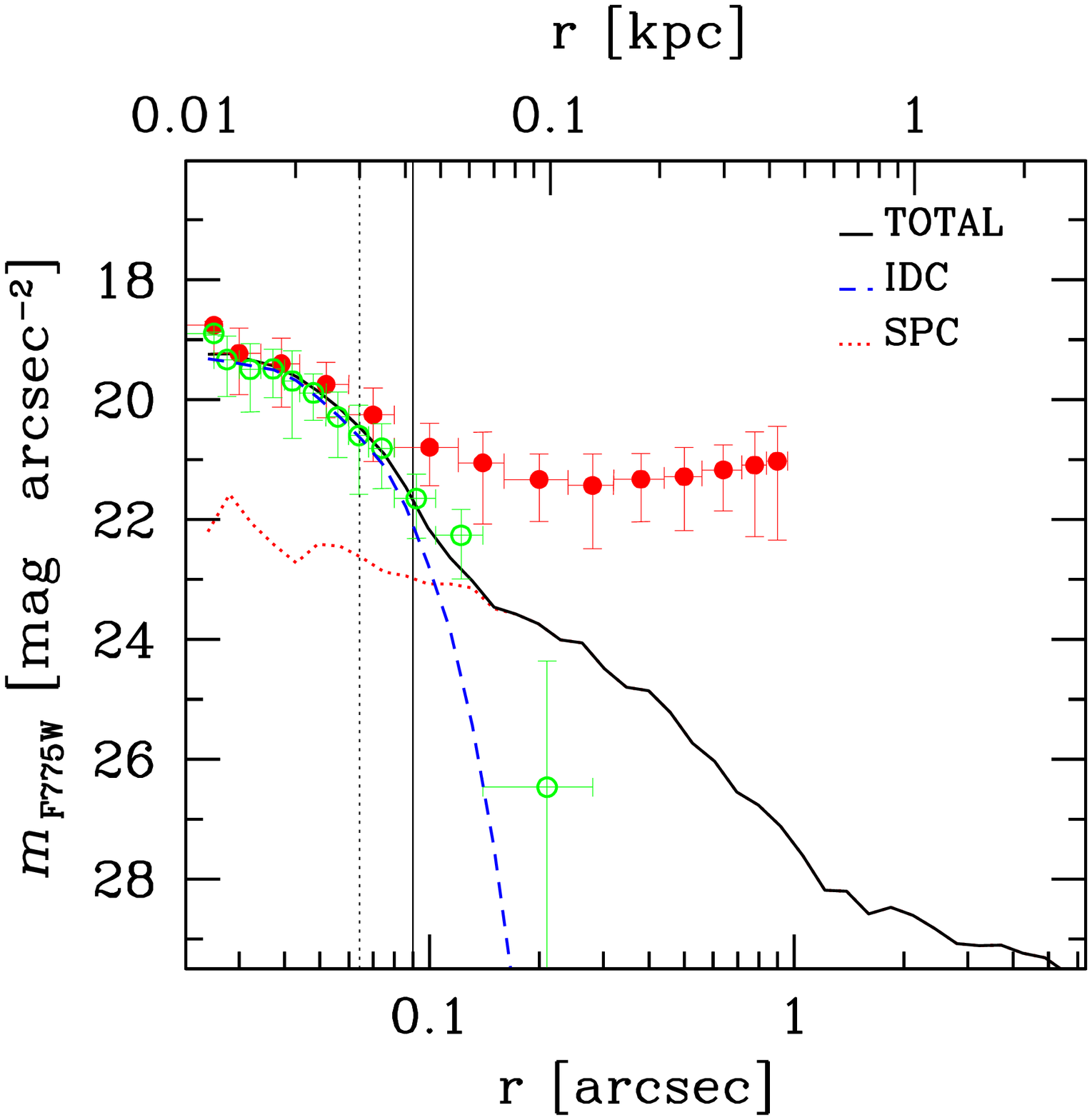,width=7.0cm} 
\epsfig{figure=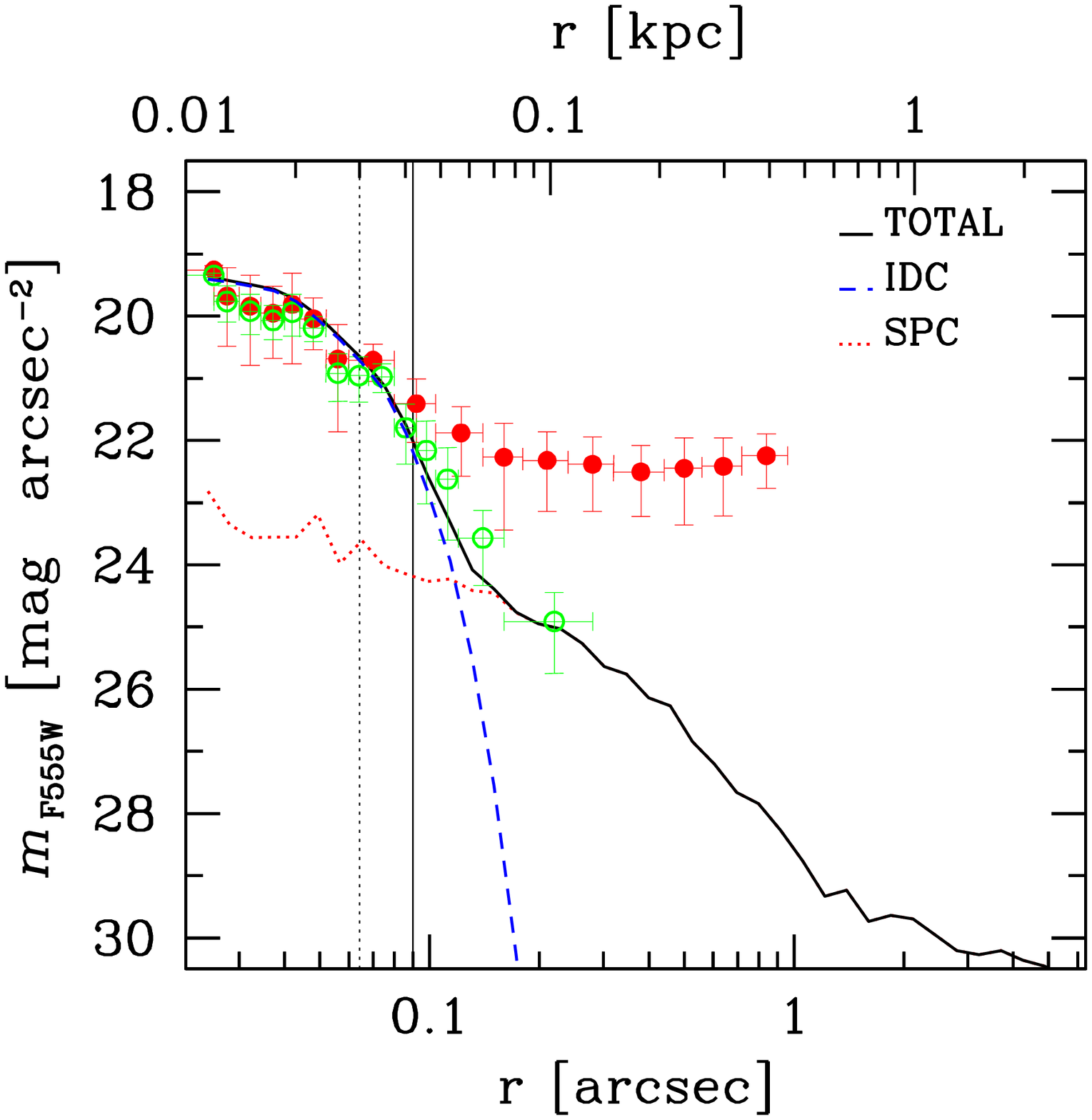,width=7.0cm} 
\epsfig{figure=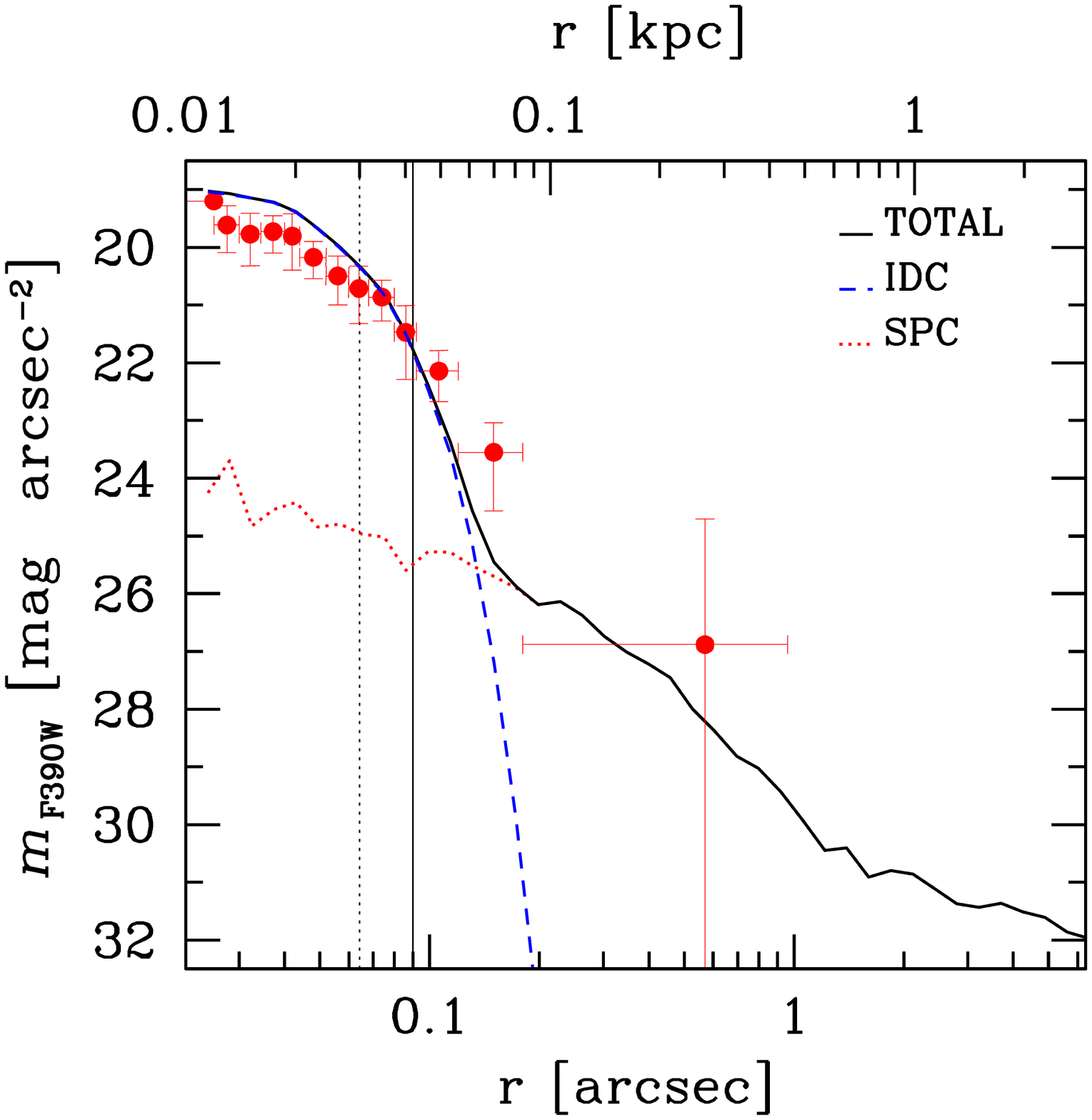,width=7.0cm}  
\epsfig{figure=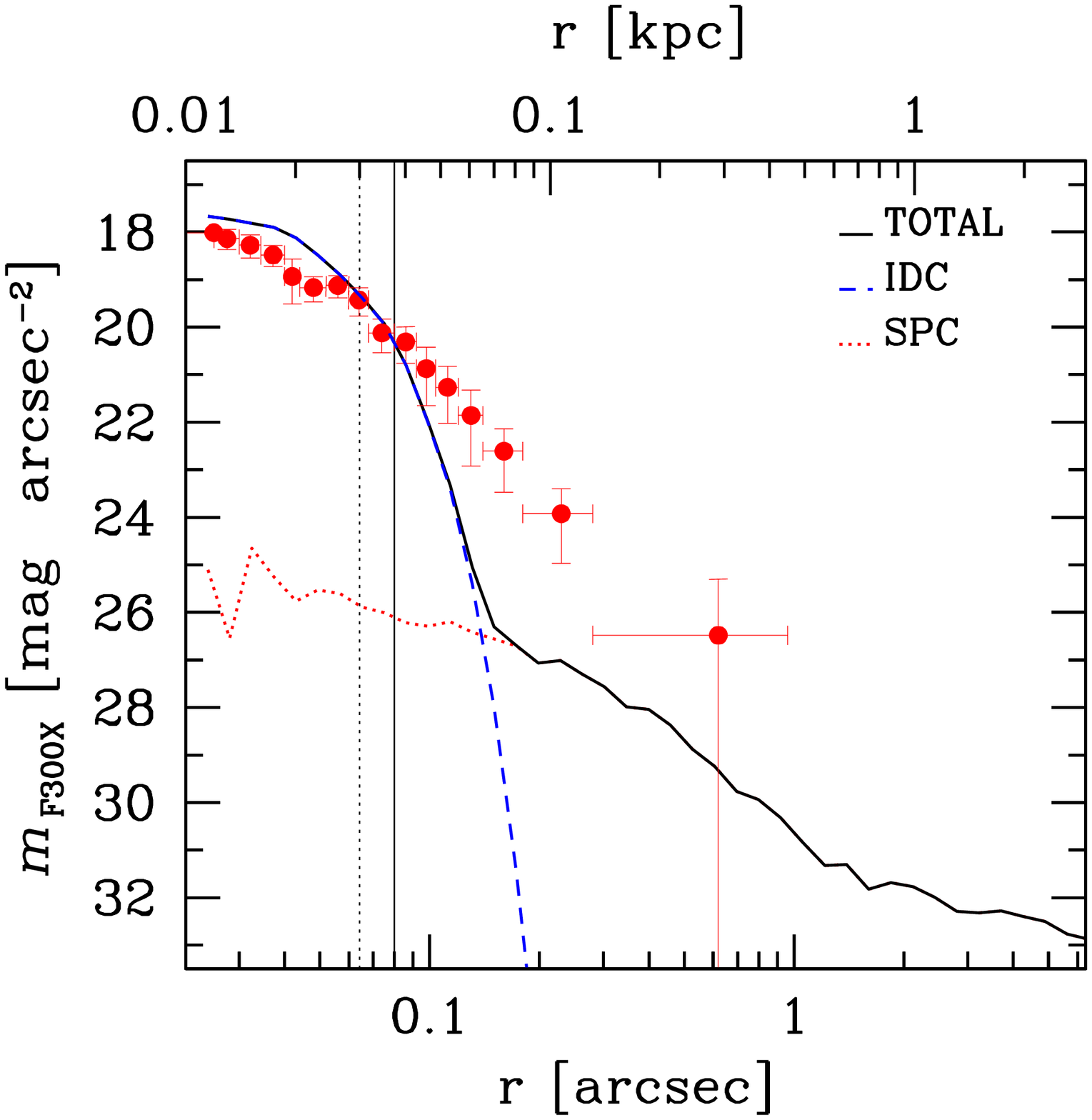,width=7.0cm} 
\epsfig{figure=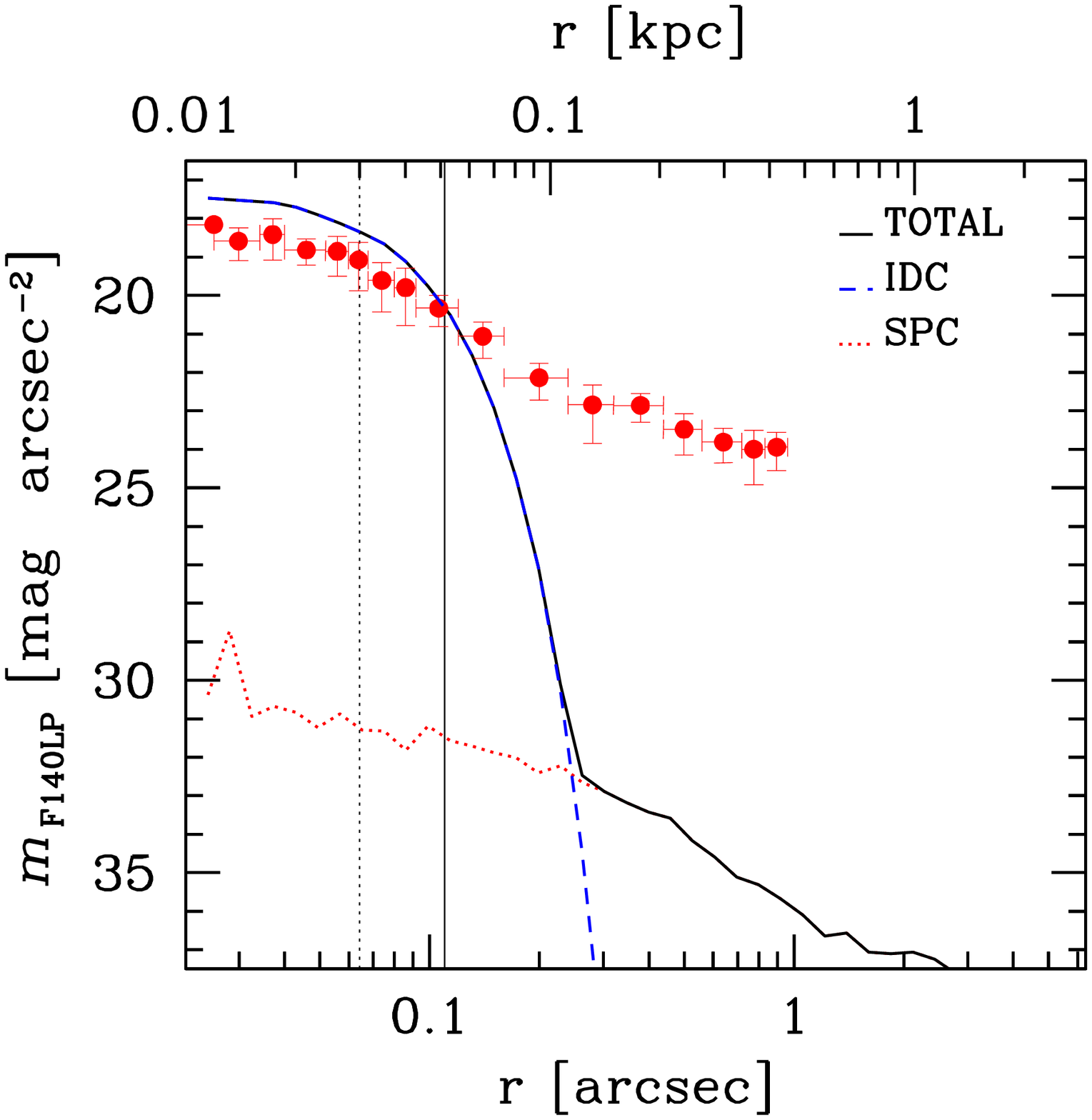,width=7.0cm}  
}}
\caption{\label{fig:fig9}
Predicted surface brightness profiles of the HLX-1 counterpart, obtained by combining the stellar population in run~C and the model of irradiated disc. 
From left to right and from top to bottom: filter F160W, F775W, F555W, F390W, F300X and F140LP. In all  panels, dotted line (red on the web):  simulated profile in run~C at $t=2.5$ Gyr after the first pericentre passage; dashed line (blue on the web): contribution of the irradiated disc component (IDC), smoothed over the PSF; solid black line: total profile from the model. We assume E(B-V)=0 for the models.
Vertical dotted line: softening length. Vertical solid line: PSF FWHM (see Table~1).
Filled red circles and open green circles: observational surface brightness profiles of the HLX-1 counterpart, derived using the first and the second approach, respectively (the same as in Fig.~\ref{fig:fig6}). The errors are at 1 $\sigma{}$.
}
\end{figure*}
From Fig.~\ref{fig:fig6} it is apparent that the stellar population associated with the disrupted satellite cannot account for a significant fraction of the light  from the HLX-1 counterpart in the bluer filters (F390W, F300X and F140LP). To match the properties of the HLX-1 counterpart, we assume a strong contribution from disc irradiation. 
 We thus combine the stellar population component (SPC), derived from our simulations, with an irradiated disc component (IDC), in the following way.

We fix the SPC corresponding to the disrupted satellite in run~C at $t=2.5$ Gyr after the first pericentre passage, assuming $Z=0.008$ (for the satellite) and a Kroupa IMF (see the corresponding line in Table~4). This case was selected, among all the simulations, because run~C is the highest resolution simulation and because this particular snapshot reproduces better the properties of the HLX-1 counterpart in the F160W, F775W and F555W filters. 

After fixing the SPC, we combine the flux  from the SPC with the flux  from an IDC, calculated in the following way. The IDC was modelled with a code developed for computing the optical luminosity of ULX binaries (Patruno \&{} Zampieri  2008, 2010).
A standard Shakura-Sunyaev disc is assumed and both the
X-ray irradiation of the companion and the self-irradiation
of the disc are accounted for. A simplified description of 
radiative transfer for the interaction of the X-rays with the disc 
and donor surfaces is adopted (an X-ray illuminated plane-parallel atmosphere 
in radiative equilibrium; e.g. Copperwheat et al. 2005). The model does not include a Compton tail and the emission from the donor star (which is expected to be small with respect to the emission from the disc). For more details
about the IDC code we refer to Patruno \&{} Zampieri (2008, 2010).

For the IDC,  we assume BH mass $m_{\rm BH}=10^4$ M$_\odot$, bolometric luminosity equal to the Eddington luminosity ($L_{\rm Edd}=1.3\times{}10^{42}$ erg s$^{-1}$), and  inner radius $r_{\rm in}=3\,{}r_{\rm g}$ (where $r_{\rm g}$ is the gravitational radius). We leave three free parameters: the outer disc radius ($r_{\rm out}$), the albedo ($alb$) and the inclination of the disc ($i$).  
 Assuming a reddening $A_{\rm V}=0.18$, we find the following best-matching values for the free parameters: $r_{\rm out}\approx{}3.4\times{}10^{13}$ cm, $alb \approx{} 0.55$ and $i \approx{} 40^{\circ{}}$. 
The resulting fraction of the flux thermalised in the outer disc ($f_{\rm th}$) is $6.67\times{}10^{-3}$. 

 The best-matching total magnitudes (inside 0.4 arcsec) in the six {\it HST} filters are shown in Table~5. In Fig.~\ref{fig:fig9}, we show the resulting surface brightness profiles for the best-matching SPC+IDC model. To obtain the surface brightness profile, the contribution of the IDC was smoothed over a two-dimensional Gaussian profile with the same FWHM as the PSF.

Fig.~\ref{fig:fig9} and Table~5 show that there is good agreement between the observed magnitude of the HLX-1 counterpart and the model that combines SPC and IDC,  in the filters F775W and F555W.  In the F160W filter, the total magnitude of the model ($=22.82$) is lower than the observed one (23.49$\pm{}0.26$ in our Table~1, or 23.34$\pm{}0.30$ from F12), but still consistent considering the large error bars. 
In the F140LP filter, the observed profile shows a tail at $r\gtrsim{}0.2$ arcsec that extends beyond the PSF limits (see the discussion in the previous sections).

 Finally, we note that adding an IDC is just one of the possible ways to account for the flux within 0.1 arcsec. Another possible solution consists in adding a compact SC with a young stellar population  (but this would imply a completely different scenario with respect to the mergers simulated in the current paper: there are no physical reasons for a young SC to form in the satellite galaxy at $t\gtrsim{}2.5$ Gyr, when all the gas was stripped away). On the other hand, the disc component is surely present in the observations, although we cannot quantify its relative contribution uniquely (as far as we do not have accurate estimates of the source optical variability). Instead, there is no definitive evidence for a young SC surrounding HLX-1. Actually, the existence of a massive young SC in a relatively passive S0 galaxy is something very peculiar, unless we assume that this is the nuclear SC of a disrupted satellite galaxy.

\section{Star formation and SED}~\label{sec:SEDbulge}
\begin{figure*}
\center{{
\epsfig{figure=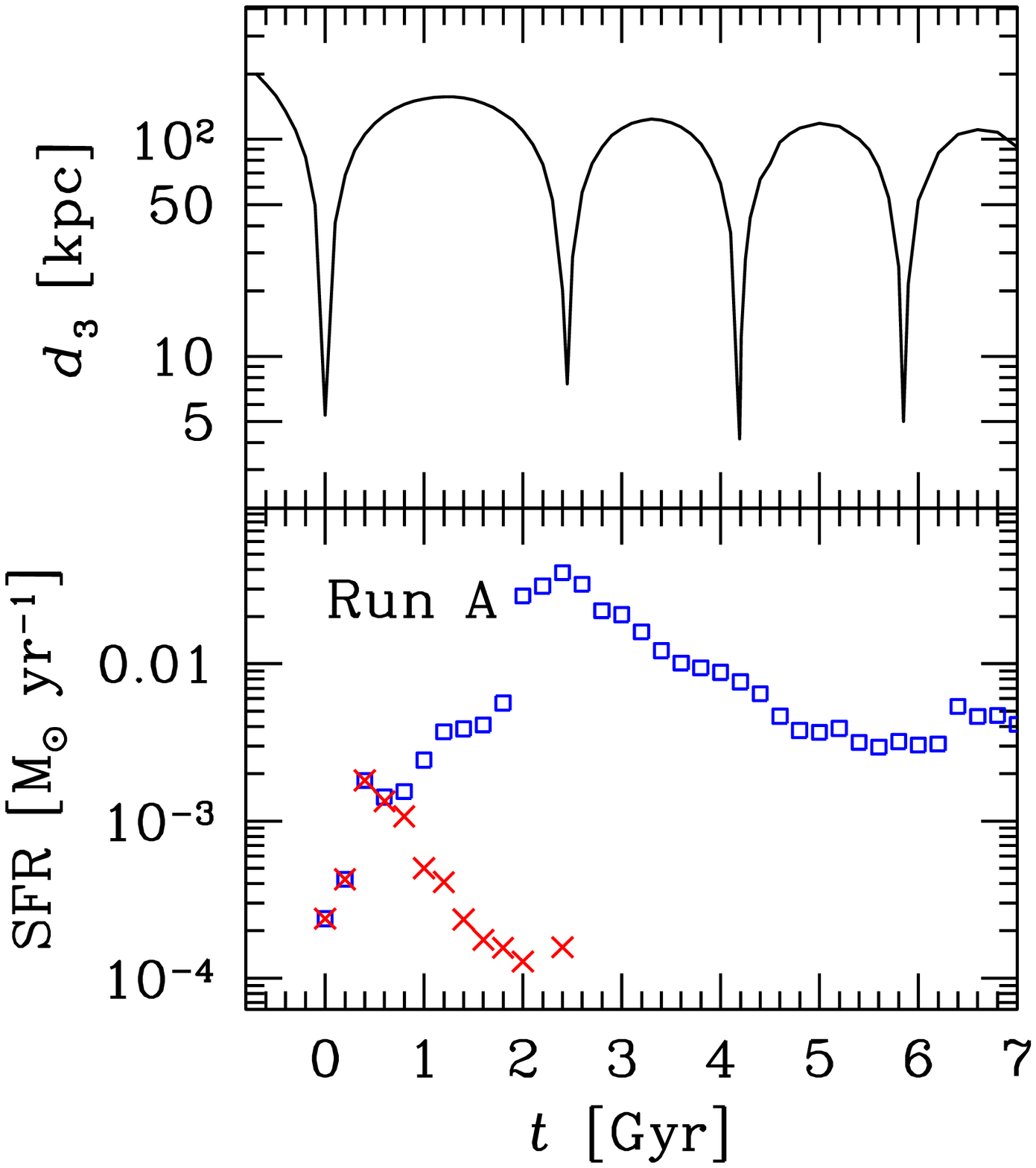,width=5.5cm} 
\epsfig{figure=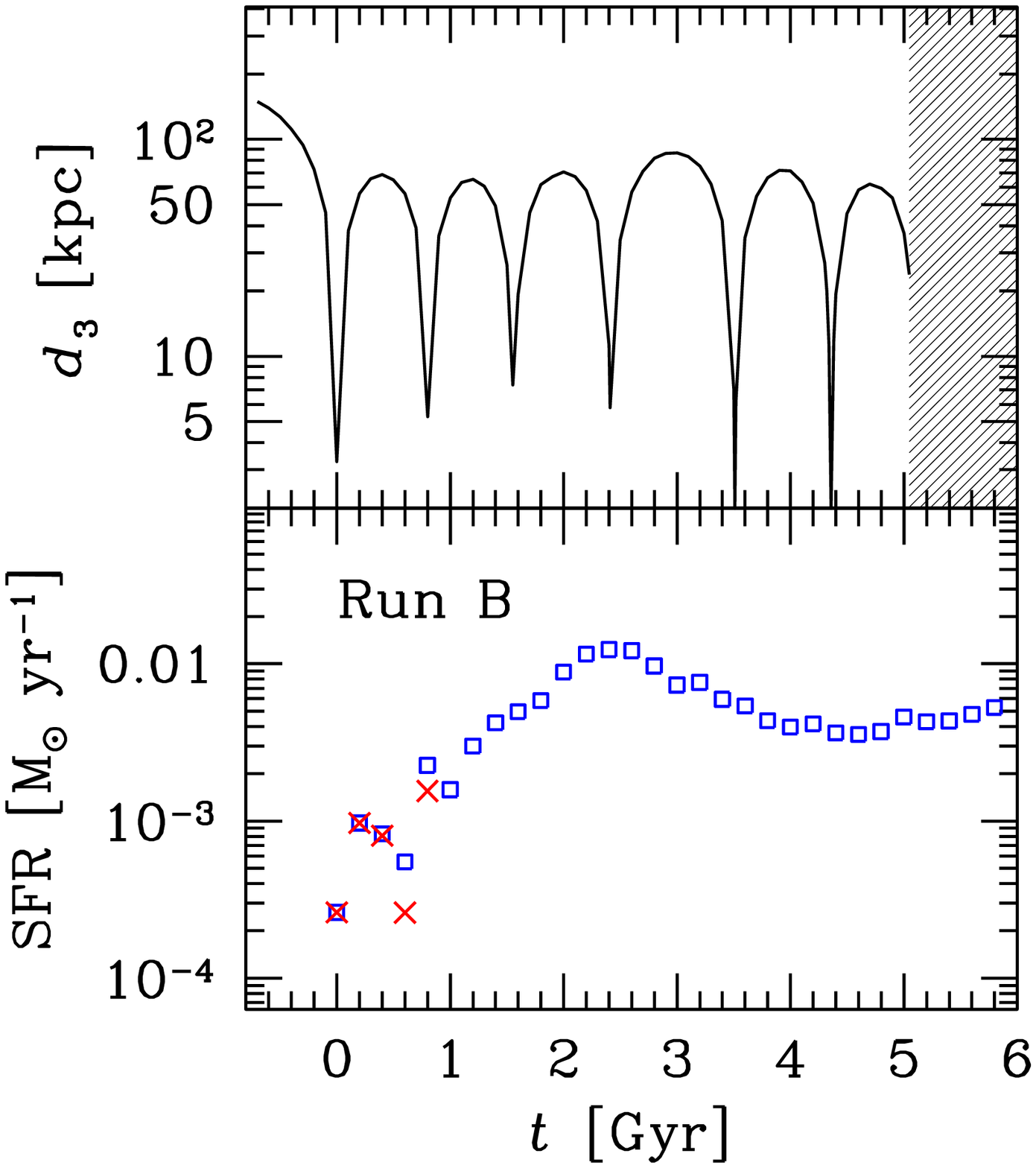,width=5.5cm} 
\epsfig{figure=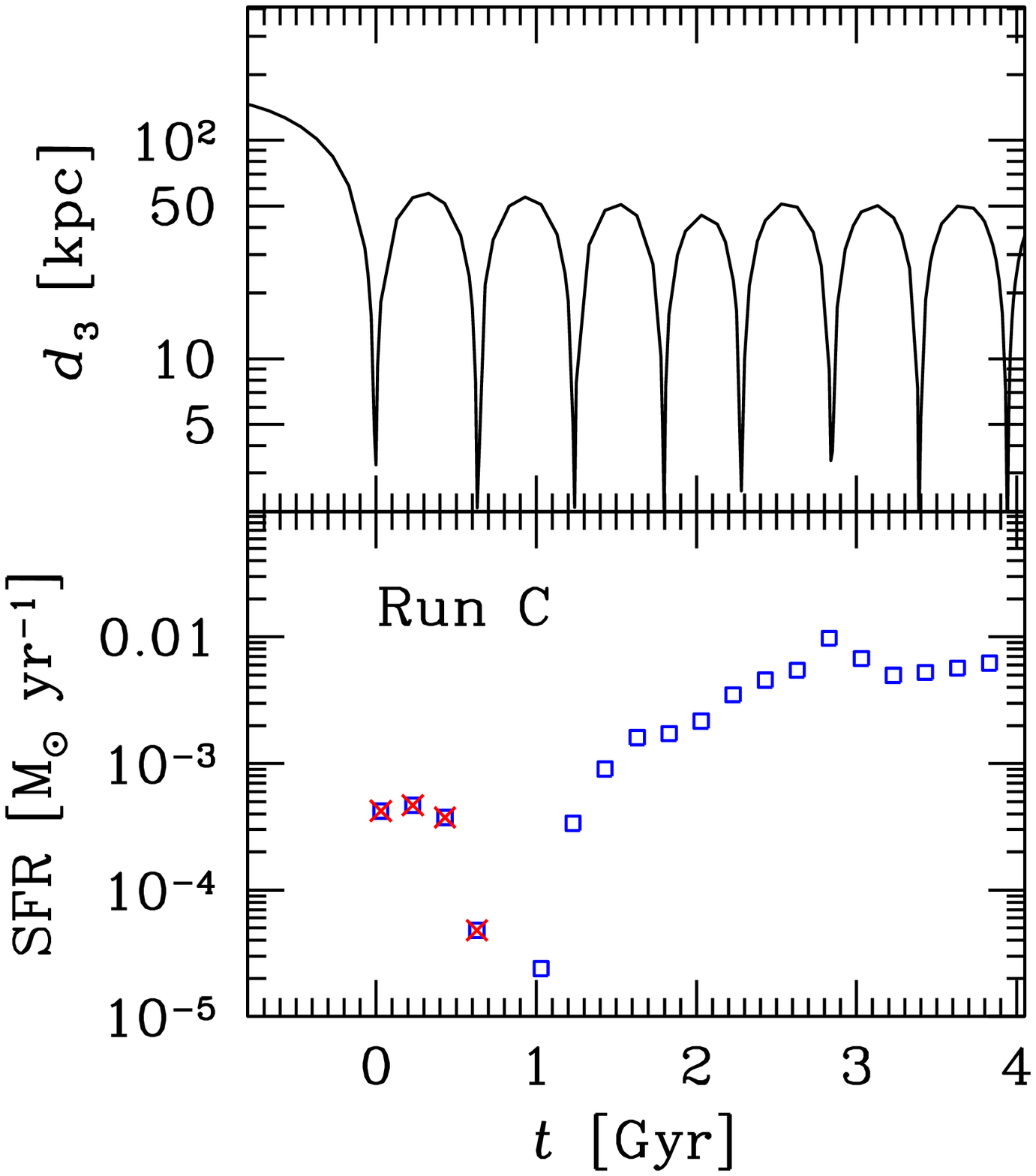,width=5.5cm} 
\epsfig{figure=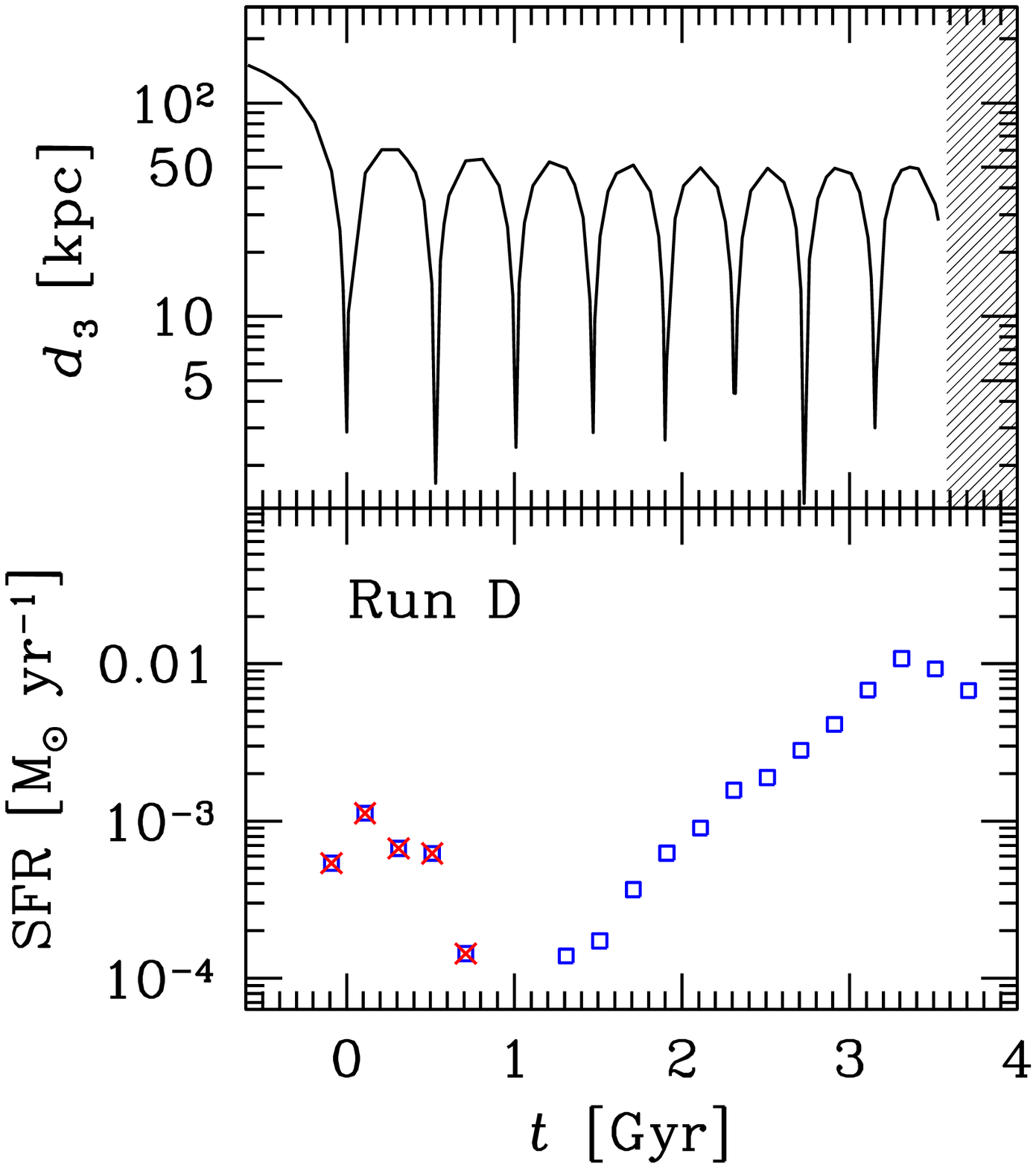,width=5.5cm}  
\epsfig{figure=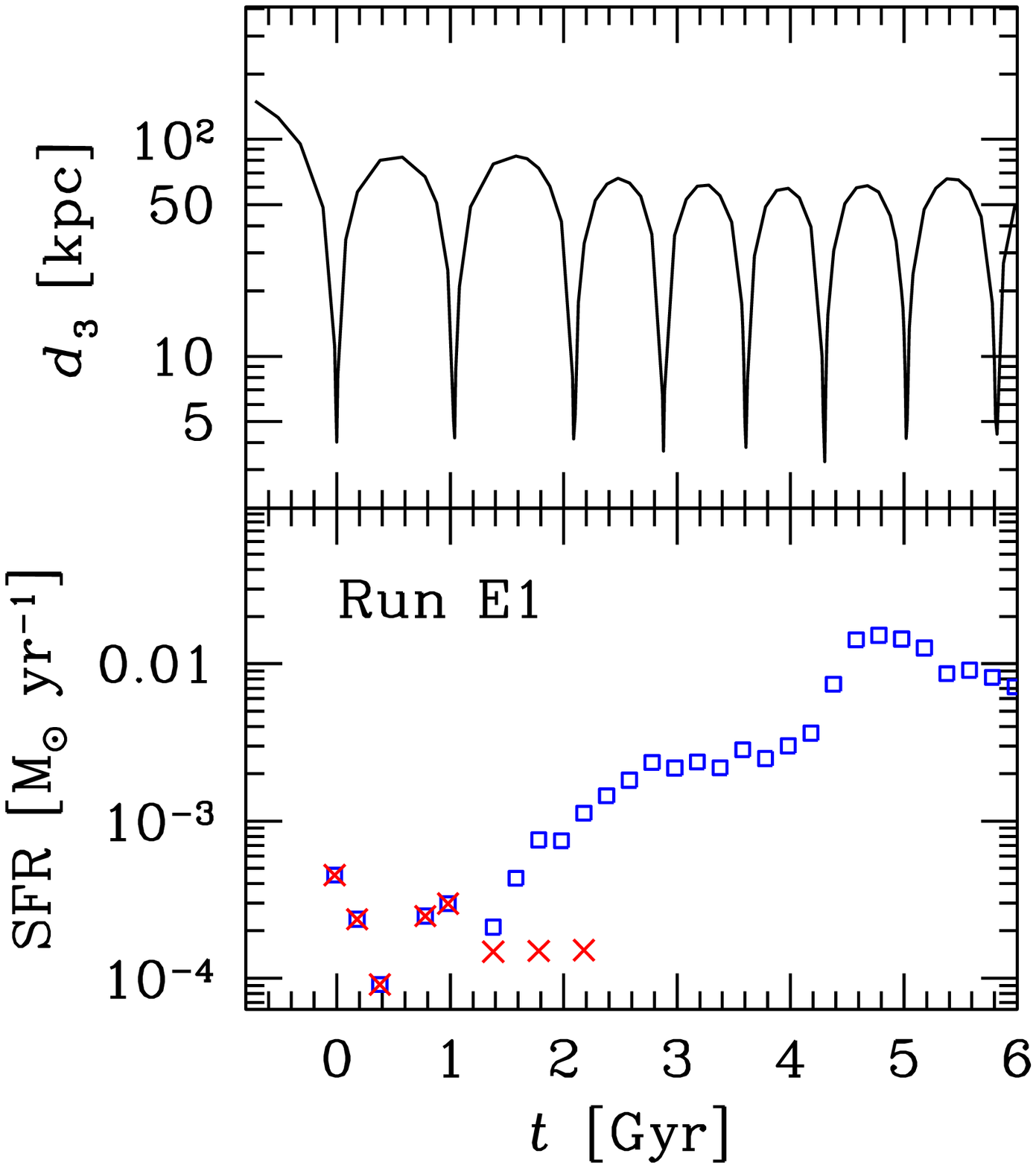,width=5.5cm}  
\epsfig{figure=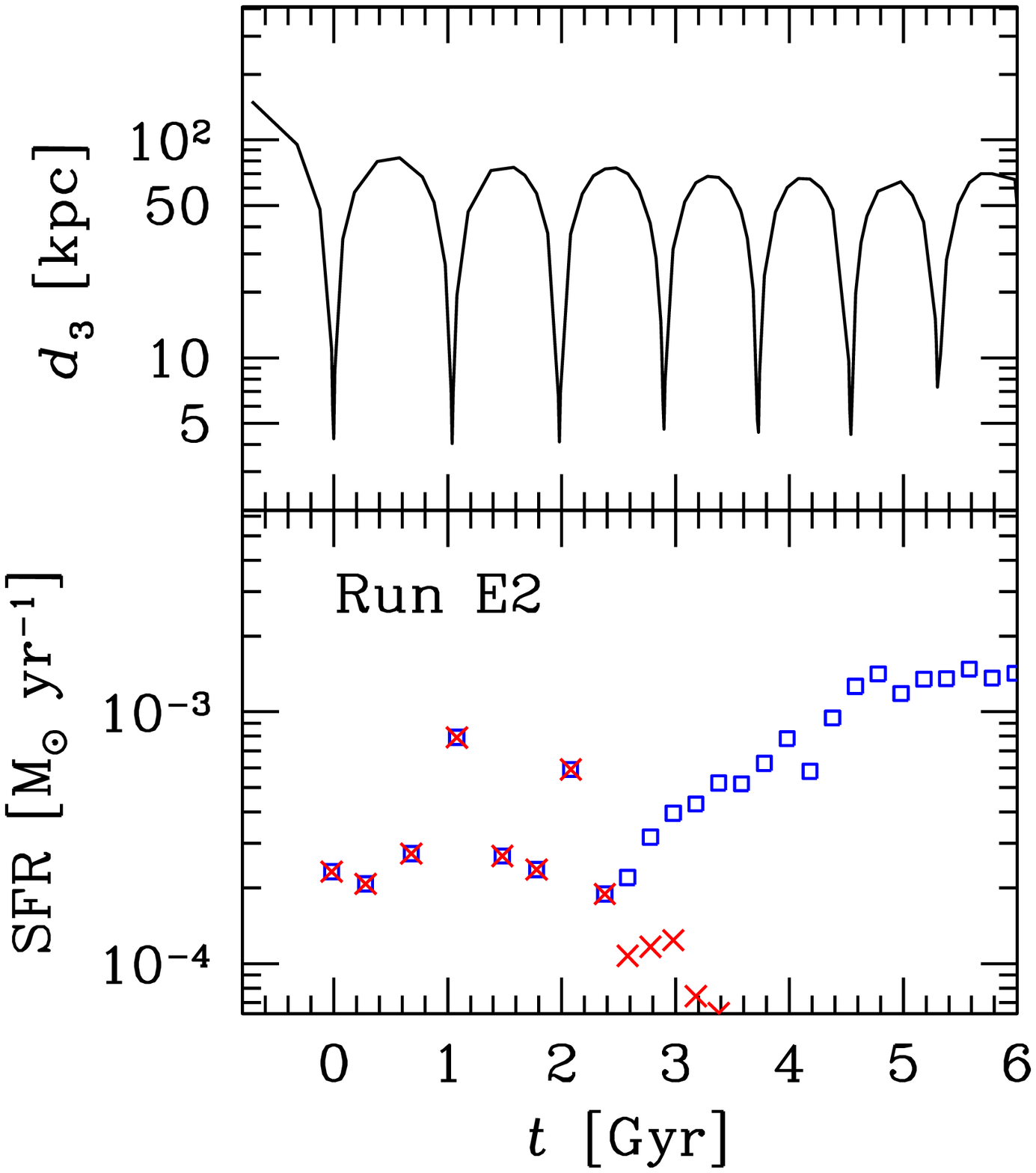,width=5.5cm}  
}}
\caption{\label{fig:fig10}
From  left to right and from top to bottom: run~A, B, C, D, E1 and E2. For each run, we show the three-dimensional distance $d_3$ between the nuclei of the two galaxies (top) and the SFR (bottom) as a function of the time elapsed since the first pericentre passage. Crosses (red on the web): SFR in the satellite; open squares (blue on the web): total SFR in the simulation. The shaded area in the panels of run~B and of run~D indicates that $d_3$ cannot be estimated anymore, as the nucleus of the satellite galaxy is no longer self-bound.
}
\end{figure*}
S10 analyze optical images from the Magellanic Telescope, UV images from the {\it Swift}/Ultraviolet Optical Telescope (UVOT), and, using stellar population modelling, find that the bright far UV emission from ESO~243-49 can be accounted for by an ongoing star formation at a rate of $\sim{}0.03$ M$_\odot{}$ yr$^{-1}$  (integrated over the entire galaxy). 
 
In this Section, we study the SF history in the S0 galaxy, combining $N-$body simulations with stellar population models, and comparing them with observations, to look for the possible link with a recent minor merger. 
\subsection{Simulated SFR}~\label{sec:SFR}
From the simulations, we can trace the SFR as a function of time since the beginning of the interaction. Fig.~\ref{fig:fig10} shows the SFR as a function of time for all the runs. In the same figures, the relative three-dimensional distance $d_3$ between the nuclei of the two galaxies is shown, to understand the orbital evolution of the satellite remnant and its connection with the SFR. Note that run~A and run~B in Fig.~\ref{fig:fig10} are the same as fig. 3 and fig. 4 of paper~I, respectively, but integrated for a longer time\footnote{Note that figs. 3 and 4 of paper~I contain a typo: the open square corresponding to the total SFR at $t=0$ is missing in both figures.}.
\begin{table}
\begin{center}
\caption{Properties of the bulge of ESO~243-49 from the {\it HST} data.}
 \leavevmode
\begin{tabular}[!h]{ccc}
\hline
Band & Filter & Flux$^{\rm a}$ \\
     &        & (10$^{-15}$ erg cm$^{-2}$ s$^{-1}$ \AA{}$^{-1}$)\\
\hline
FUV & F140LP & $0.13\pm{}0.01$ \\ 
NUV & F300X  & $0.22\pm{}0.02$ \\ 
C   & F390W  & $1.06\pm{}0.12$ \\ 
V   & F555W  & $2.64\pm{}0.29$ \\ 
I   & F775W  & $2.89\pm{}0.27$ \\ 
H   & F160W  & $1.87\pm{}0.33$ \\ 
\noalign{\vspace{0.1cm}}
\hline
\end{tabular}
\begin{flushleft}
\footnotesize{$^{\rm a}$ Observed flux within 3\farcs1 ($\sim{}1.43$ kpc) from the centre of ESO~243-49. The errors are at 1 $\sigma$.}
\end{flushleft}
\end{center}
\end{table}

As already noted in paper I, the SFR is triggered by the first pericentre passage and is initially concentrated in the satellite, because the gas is not yet stripped. After a few pericentre passages ($N_{\rm p}\sim{}2-4$), the gas is almost totally stripped from the satellite, where the SF stops.  The SF in the bulge of the S0 starts as soon as the stripped gas flows toward the centre of the primary galaxy. In the bulge, the SF goes on for the entire duration of the merger (and likely even for longer times, as evident from the panels of  Fig.~\ref{fig:fig10} corresponding to runs B and D), with a SFR ranging from a few $\times{}10^{-3}$ M$_\odot$ yr$^{-1}$ up to a few  $\times{}10^{-2}$ M$_\odot$ yr$^{-1}$ (but this value likely depends on the initial gas content of the satellite). The comparison between run~E1 (prograde) and run~E2 (retrograde) indicates that the SFR may be significantly higher if the orbit is prograde.
\begin{figure}
\center{{
\epsfig{figure=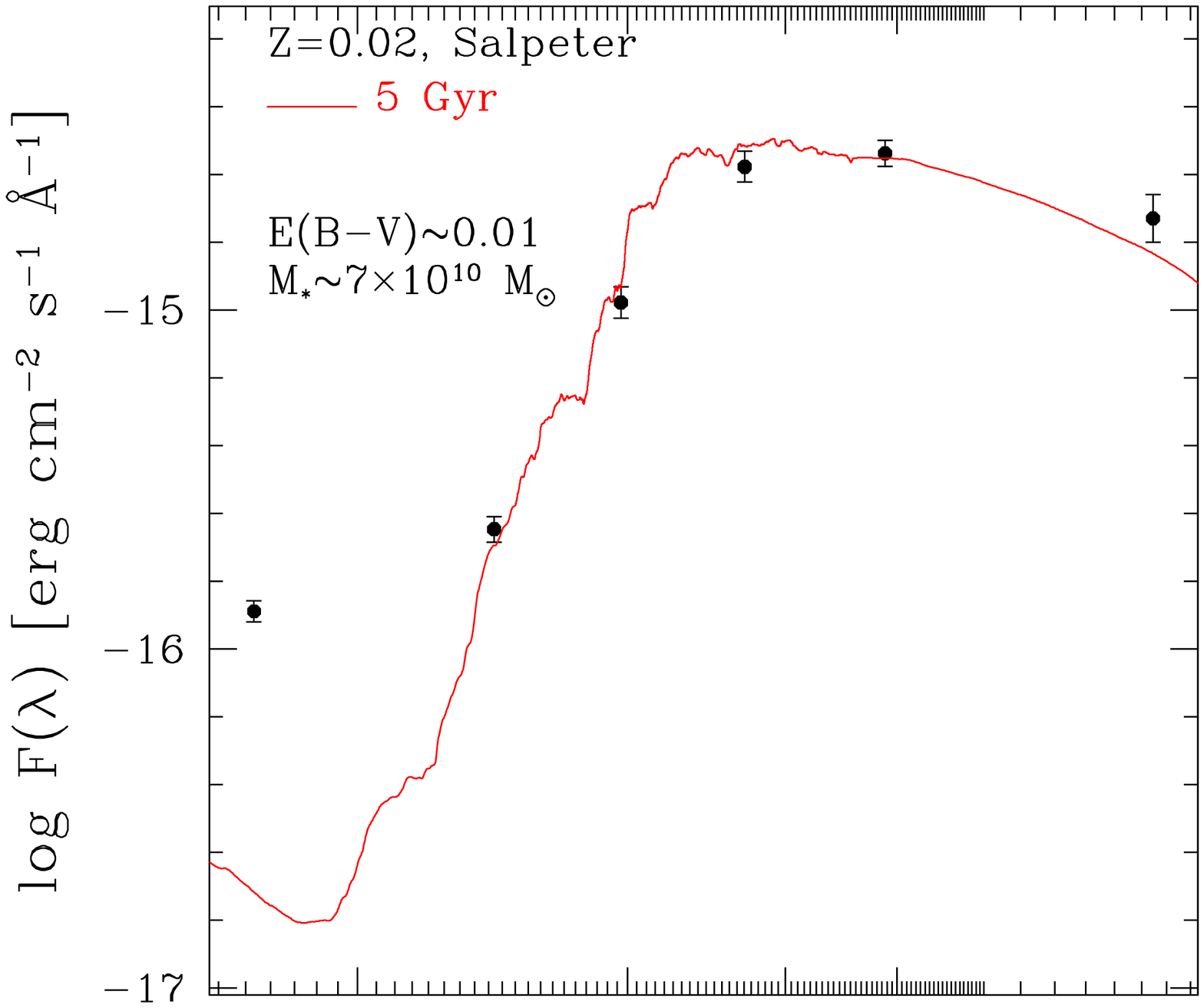,width=7cm} 
\epsfig{figure=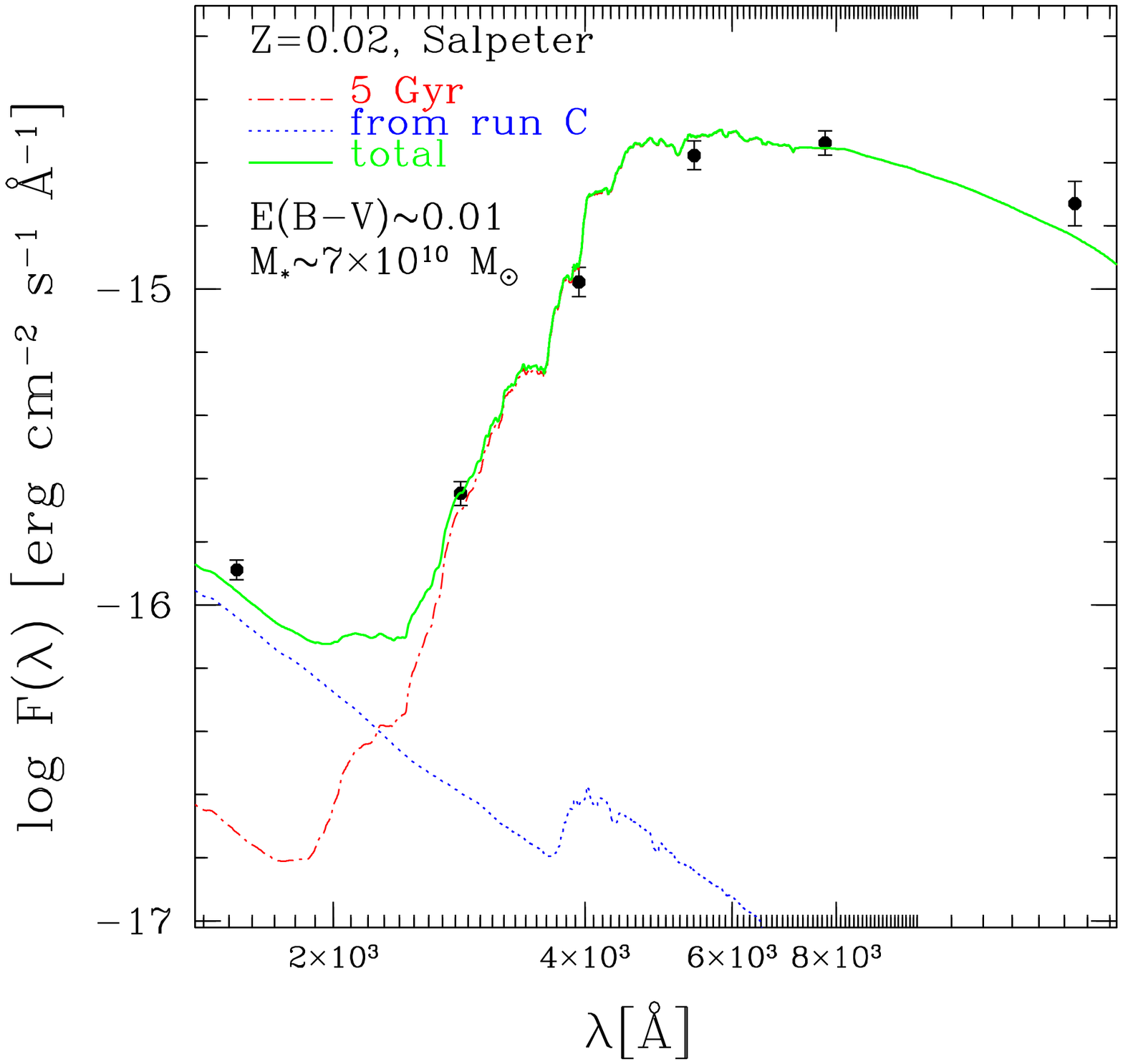,width=7cm} 
}}
\caption{\label{fig:fig11}
Observed SED of the bulge of ESO~243-49 (filled circles) compared with the best-matching stellar population models (assuming a Salpeter IMF). The errors are at 1 $\sigma{}$.
In the top panel, solid line (red on the web):  model with a single old stellar population (SSP+Sa). Bottom panel:  model with a old stellar population plus a young stellar population, adopting the SF history of run~C (run~C+Sa). Dotted line (blue on the web): young stellar population; dot-dashed line (red on the web): old stellar population; solid line (green on the web): total flux in the model. 
}
\end{figure}
\subsection{Observed SED of the bulge of ESO~243-49}~\label{sec:bulge}
We select a circle of radius 3\farcs1  ($\sim{}1.43$ kpc) centred on the bulge of ESO~243-49, corresponding to the central regions of the bulge. The corresponding fluxes in the six different filters, obtained with the {\it phot} task in IRAF, are listed in Table~6. We exclude the outer parts of the bulge to minimize contamination by the BGG in the FUV. 
To check whether the BGG can contaminate the bulge significantly, we divided the selected circle into four quadrants,  
 and we extracted the flux in the six filters for each quadrant. As $\lesssim{}10$ per cent relative variations were found from one quadrant to another (and in particular between the FUV filter and the other filters), we conclude that the contamination from the BGG is not important in the selected area.

 In order to model the observed SED of the bulge of ESO 243-49, we use again the SSP models based on the tracks of Marigo et al. (2008, see Section~\ref{sec:simul} for details). We assume a metallicity $Z=0.02$ (i.e. slightly supra-solar metallicity) and  an age $t_{\rm old}=5$ Gyr for the old stellar population of ESO~243-49, consistent with the estimate of $t_{\rm old}=4.5^{+4}_{-2.5}$ Gyr derived in S10 (based on the observed strength of the H$\beta{}$ absorption line, and of the Fe5270 and Fe5335 indexes). We adopt a minimum reddening parameter E(B-V) $=0.01$, which is the foreground value for ESO~243-49, derived by the NASA/IPAC Extragalactic Database (NED). 
 Finally, we use either a Salpeter or a Kroupa IMF.


In the first two lines of Table~7, we report the properties of the two models, named SSP+Sa and SSP+Kr, that assume that there is a single old stellar population, with instantaneous SF and with total mass $M_{\rm old}$. SSP+Sa and SSP+Kr provide the best match for a Salpeter and for a Kroupa IMF, respectively. The SED associated with  SSP+Sa is shown in the top panel of Fig.~\ref{fig:fig11}. Here, we have plotted the SSPs kindly made available by A. Bressan (see also Chavez et al. 2009), based on the Padova models (Bertelli et al. 1994; Bressan et al. 1998) and on the MILES spectral library (Sanchez-Blazquez et al. 2006), to  visualize the whole spectrum. 

SSP+Sa and  SSP+Kr differ for $M_{\rm old}$, which is a factor of three lower for a Kroupa IMF. They match the data, except for the FUV: the observed flux in the FUV is a factor of $\sim{}10$ higher than predicted by the models. Increasing the reddening parameter E(B-V) makes this discrepancy even worse, as it increases the difference between infrared and FUV filters.
This is the main evidence for the existence of a younger stellar population in the bulge of ESO~243-49.

\begin{table*}
\begin{center}
\caption{Best matching stellar population models for the bulge of ESO~243-49.}
 \leavevmode
\begin{tabular}[!h]{ccccccccccc}
\hline
Model & $M_{\rm old}$ & $t_{\rm old}$ & $f_{\rm y}$ & E(B-V) & F140LP &  F300X & F390W & F555W & F775W & F160W \\
&  (10$^{10}$ M$_\odot{}$) & (Gyr) & &  & \multicolumn{6}{c}{(10$^{-15}$ erg cm$^{-2}$ s$^{-1}$ \AA{}$^{-1}$)} \\
\hline
SSP+Sa  & 7.0 & 5 & 0                    & 0.01 & 1.1$\times{}10^{-2}$ & 0.35 & 1.34 & 2.74 & 2.63 & 1.57   \\
SSP+Kr  & 2.4 & 5 & 0                    & 0.01 & 1.1$\times{}10^{-2}$ & 0.34 & 1.30 & 2.66 & 2.80 & 1.51   \\
Run~C+Sa & 7.0 & 5 & 3.8$\times{}10^{-4}$ & 0.01 & 7.7$\times{}10^{-2}$ & 0.37 & 1.34 & 2.74 & 2.62 & 1.56    \\
Run~C+Kr & 2.4 & 5 & 3.8$\times{}10^{-4}$ & 0.01 & 8.4$\times{}10^{-2}$ & 0.36 & 1.32 & 2.67 & 2.81 & 1.51    \\
\noalign{\vspace{0.1cm}}
\hline
\end{tabular}
\begin{flushleft}
\footnotesize{SSP+Sa: best-matching model with an old single stellar population and Salpeter IMF; SSP+Kr: the same as SSP+Sa but for a Kroupa IMF; 
Run~C+Sa: model  with an old plus a young stellar population and Salpeter IMF. In this model, we assume that the young stellar population follows the SF history of run~C. Run~C+Kr: the same as  Run~C+Sa but for a Kroupa IMF.\\$M_{\rm old}$ and $t_{\rm old}$  are the mass and the age of the old stellar component, respectively. $f_{\rm y}$ is the mass fraction of the young stellar component with respect to the old stellar component. E(B-V) is the best-matching reddening parameter. F140LP,  F300X, F390W, F555W, F775W and F160W are the fluxes associated with the F140LP,  F300X, F390W, F555W, F775W and F160W filters, respectively, and are all expressed in units of 10$^{-15}$ erg cm$^{-2}$ s$^{-1}$ \AA{}$^{-1}$.} 
\end{flushleft}
\end{center}
\end{table*}

Then, we ran a model where the SFR and the SF history are fixed and taken from our run~C. In particular, we assume that  run~C at $t=3$ Gyr after the first pericentre passage represents the present time. This is justified by the fact that, in run~C, the stellar population in the satellite remnant  at $t=3$ Gyr is consistent with the properties of the counterpart of HLX-1 (see Table~4 and Fig.~\ref{fig:fig6}). We assume metallicity $Z=0.02$ and a Kroupa IMF. The Kroupa IMF is a natural choice for comparison with the $N-$body simulation, by construction. In fact, in the initial conditions, we assumed that the total initial stellar mass of the S0 galaxy is $7\times{}10^{10}$ M$_\odot$ to match the value obtained from the SED by S10, and based on a  Kroupa IMF (see Section~\ref{sec:nbody}).

We model the SED corresponding to an old population with $t_{\rm old}=5$ Gyr and $M_{\rm old}=2.4\times{}10^{10}$ M$_\odot$ (corresponding to the stellar mass enclosed within a radius of 1.4 kpc centred on the bulge of the S0 galaxy in run~C), plus a young stellar population whose mass and SF history are completely determined in a self-consistent way by run~C (as shown in Fig.~\ref{fig:fig10}). 
We can repeat the same analysis also for a Salpeter IMF, but in this case we have to take a three times larger value for $M_{\rm old}$ ($=7.0\times{}10^{10}$ M$_\odot$), to obtain similar results.

The results are named run~C+Sa and run~C+Kr (see Table~7 and the bottom panel of Fig.~\ref{fig:fig11}). These models match quite well the observed SED, especially because masses and ages of the stellar populations are fixed, being completely determined by the $N-$body simulation. 
The main difference between the SED obtained from run~C and the observed SED is in the FUV, where the simulation accounts only for $\sim{}60$ per cent of the observed flux (increasing the reddening parameter E(B-V) makes this discrepancy worse). 
On the other hand, we note that the method to implement the SF in most SPH codes is a stochastic approach and cannot capture the microphysics details of SF. Even if the code we adopt includes one of the most accurate methods for implementing the SF, a factor of two difference with respect to the observed SFR is well within the uncertainty of the adopted technique.

\section{Discussion on the merger scenario}~\label{sec:discussion}
 The aim of this paper is to compare the main observational features of the HLX-1 counterpart with the properties predicted from the simulations, assuming that the HLX-1 counterpart is the nucleus of a disrupted satellite galaxy. Finding observational evidence for a recent minor-merger involving ESO~243-49 is thus beyond our aims. 
However, in this Section we briefly discuss the observational arguments in favour and against the minor-merger scenario, and we speculate which are the alternative explanations. The scope of this section is to propose which observations may effectively shed light on the nature of ESO~243-49 and of the HLX-1 counterpart.

As we summarized in the introduction, the hypothesis that  ESO~243-49 recently underwent a minor merger is indirectly suggested by (i) the presence of prominent dust lanes around its nucleus  (Finkelman et al. 2010; Kaviraj et al. 2012; Shabala et al. 2012) and (ii) the evidence of UV emission centred on its bulge (S10), indicating ongoing star formation (SF, Kaviraj et al. 2009, 2011; paper~I).

\subsection{Dust lanes in S0 galaxies}
The presence of dust lanes in S0 galaxies is not an indisputable evidence for recent minor mergers. On the other hand, recent theoretical and observational work (e.g. Lauer et al. 2005; Sarzi et al. 2006; Sim$\tilde{o}$es Lopes et al. 2007; Schawinski et al. 2007; Zhang, Gu \&{} Ho 2008; Schawinski et al.  2009;  Shabala et al. 2012) supports the idea that internal sources of dust (e.g. stellar winds) are not sufficient to explain dust lanes in early-type galaxies. On the contrary, there are various reasons to think that a large fraction of this dust has external origin. Firstly, internally created dust should be more uniformly distributed than the observed dust, particularly in lenticular galaxies (e.g., Sim$\tilde{o}$es Lopes et al. 2007). Secondly, internal dust creation via stellar mass loss appears to be inconsistent with the absence of dust in approximately 50 per cent of all early-type galaxies  (e.g., Sim$\tilde{o}$es Lopes et al. 2007), unless we assume that some event destroyed/removed all the missing dust. 
Thirdly, the kinematics of gas and dust is often decoupled from that of stars, suggesting an external origin. For example, counter-rotating gaseous discs were observed in a number of S0 galaxies (Bertola, Buson \&{} Zeilinger 1992; Sarzi et al. 2006; Morganti et al. 2006). Furthermore, the ionized gas component is kinematically misaligned with respect to the stars in $36\pm{}5$ per cent of the fast rotating early-type galaxies in the ATLAS$^{\rm 3D}$ survey (e.g. Davis et al. 2011, 2013). This percentage is even higher for slow rotators. As ionized, atomic and molecular gas are always kinematically aligned in the ATLAS$^{\rm 3D}$ survey, this result points toward a common origin for the gas and dust component, in different phases. 

Minor mergers of small gas-rich satellites and  accretion of cold and/or warm intergalactic medium are the main possible external sources of gas and dust (e.g. Sim$\tilde{o}$es Lopes et al. 2007; Shabala et al. 2012). Smooth accretion of cold and/or warm intergalactic medium is expected to be quenched at low redshift ($z<1$, e.g. Murante et al. 2012). Thus, a recent minor merger scenario is one of the most likely explanations for the dust in ESO~243-49.

\subsection{SFR in ESO~243-49 and in other S0 galaxies}
The UV data indicate a SFR $\sim{}0.03$ M$_{\odot}$ yr$^{-1}$ for ESO~243-49 (integrated over the entire galaxy).  It is possible to infer the SFR of ESO~243-49 also from the radio observations of the Phoenix Deep Survey (PDS, made at 1.4 GHz, with the Australia Telescope Compact Array, Hopkins et al. 2003). Hopkins et al. (2000) report a 1.4 GHz luminosity $L_{\rm 1.4\,{}GHz}=1.86\times{}10^{20}$ W Hz$^{-1}$ for ESO~243-49, corresponding to a ${\rm SFR}\sim{}0.25$ M$_\odot{}$ yr$^{-1}$ (assuming the conversion reported in equation 6 of Bell 2003).  This value is a factor of 10 higher than the one derived from the UV. Actually, there is a factor of 2 uncertainty in the application of the SFR$-L_{\rm 1.4\,{}GHz}$ calibration on a galaxy-by-galaxy basis (see the discussion in Bell 2003). The uncertainty in the conversion is higher (a factor of 5 or more) for both very high and very low luminosity galaxies, and the radio luminosity of ESO~243-49 is close to the low-luminosity end of the Bell (2003) sample. These {\it caveats} prevent us from claiming a discrepancy between the UV and the radio estimate of the SFR.

The optical spectrum of ESO~243-49 does not show emission lines (Hopkins et al. 2000; Jones et al. 2004, 2009; S10), which might be interpreted as lack of star forming regions. On the other hand, many S0 galaxies with UV excess have no or weak emission lines, consistent with strong absorption by local dust and/or with a lack of OB stars (Temi, Brighenti \&{} Mathews 2009). On the basis of the lack of emission lines in the optical spectrum,  Hopkins et al. (2000) argued that the radio emission of ESO~243-49 is connected with an AGN. This is difficult to reconcile with the non-detection of nuclear X-ray sources by {\it Chandra} (Servillat et al. 2011), although the presence of a heavily obscured AGN (supported by the strong dust lanes in ESO~243-49) is still consistent with the observations.

Hopkins et al. (2000) report the radio continuum for 15 galaxies in Abell~2877, out of the 70  cluster galaxies falling within the PDS area. The remaining 55 galaxies were not detected in the PDS, implying 1.4 GHz flux density $<0.1$ mJy. Thus, ESO~243-49 is one of the 15 brightest radio-continuum sources in the PDS sample of Abell~2877. Among these 15 galaxies, there are eight S0 galaxies, including ESO~243-49 (see table 1 of Hopkins et al. 2000). ESO~243-49 is the fifth brightest source at 1.4 GHz among these eight S0 galaxies. For most of these sources (even for the three S0 galaxies with emission-line spectra) a part of the radio emission may be due to an AGN.

A large fraction of early type galaxies ($\approx{}30$ per cent, Yi et al. 2005; Kaviraj et al. 2007; see Yi 2008 for a review) have an excess of UV emission, consistent with ongoing SF. The SFR rate of ESO~243-49, as derived from the UV measurements, from the analysis of the SED and even from the radio continuum, is consistent with the level of SFR estimated for other lenticular galaxies. For example, 10 out of 14 S0 galaxies observed with both SAURON and Spitzer show SFR$\gtrsim{}0.02$  M$_\odot{}$ yr$^{-1}$ (which is the threshold for SAURON detection) and $\lesssim{}0.2$  M$_\odot{}$ yr$^{-1}$ (Temi et al. 2009).

The proposed explanations for the origin of the SF in S0 galaxies (see e.g.  Salim \&{} Rich 2010; Marino et al. 2011; Salim et al. 2012) involve either mechanisms that quench the SF in a former spiral galaxy, transforming it into a S0 galaxy (e.g. the fading of the original SF in a former spiral galaxy, the quenching of the SF by AGN feedback, the removal of gas by ram-pressure and/or harassment), or mechanisms that `rejuvenate' an already passive galaxy (e.g. recycling of gas ejected by stellar winds and supernovae, smooth gas-accretion from cold/warm filaments, minor mergers). Recently, Salim et al. (2012) find that $\approx{}55$, $\approx{}25$ and $\approx{}20$  per cent of their sample of star-forming S0 galaxies are consistent with the smooth gas accretion, the fading of SF and the minor merger scenario, respectively.

Among these mechanisms, (i) smooth gas-accretion from cold/warm filaments (e.g. Martig et al. 2009;  Salim et al. 2012),  (ii) minor mergers with gas-rich satellites  (e.g. Kaviraj et al. 2009, 2011; Salim et al. 2012) and (iii) gas removal by ram pressure and/or harassment are likely efficient in galaxy clusters.
 The first systematic study of the warm gas ($10^{4-5}$ K) distribution across a galaxy cluster, namely the Virgo cluster (Yoon et al. 2012), highlights the presence of filamentary structures of warm gas, especially in the outskirts of the cluster. Such filaments are consistent with predictions from cosmological hydrodynamical simulations (Yoon et al. 2012). It is unclear whether cold gas from these filaments can be accreted by cluster galaxies before being heated up and ionized by the hot intra-cluster medium and by the hot galaxy halos.
Minor mergers are also expected to be frequent in galaxy clusters. They involve especially members of galaxy groups that recently fell into the cluster, which have still low relative velocity (Heiderman et al. 2009).
Furthermore, it is well established  that galaxies in clusters suffer from the removal of (a fraction of) their cold gas by ram pressure (e.g. Giovanelli \&{} Haynes 1985) or galaxy harassment (Moore et al. 1996). These processes can turn a spiral galaxy into a S0 galaxy, by quenching SF.

The available data about ESO~243-49 do not allow to indisputably distinguish between the minor merger, the cold gas accretion and even the gas stripping scenario. Finding a tail of ionized gas would be a strong support for the gas stripping scenario. Finding kinematic misalignment between the old and the young stellar  component would favour both the minor-merger and the gas-accretion scenarios. Finding a cold-gas filament pointing toward ESO~243-49 would be a strong evidence for the gas-accretion scenario. Unfortunately, the distance of ESO~243-49 makes difficult to observe the expected gas component.

Finally, Soria et al.  (2013) indicate a high relative velocity between ESO~243-49 and the counterpart of HLX-1 ($\sim{}420$ km s$^{-1}$, close to the escape velocity).  If the counterpart of HLX-1 was a young SC, we would expect it to move with a velocity closer to the circular velocity of nearby disc stars. Instead, most satellite galaxies move with a velocity close to the escape velocity from the primary galaxy (e.g. Khochfar \&{} Burkert 2006). Thus, the confirmation of a high velocity offset between ESO~243-49 and the counterpart of HLX-1 would be  a support for the merger scenario. Therefore, it is crucial to collect further information about the kinematics of the HLX-1 counterpart.

\section{Conclusions}~\label{sec:conclude}
The hyperluminous X-ray source HLX-1, associated with the S0 galaxy ESO~243-49 in the cluster Abell~2877, is the brightest ULX observed so far 
 and is powered by one of the strongest IMBH candidates. HLX-1 is offset by $\approx{}3$ kpc with respect to the bulge  and lies $\approx{}1$ kpc above the plane of the disc of ESO~243-49. The nature of the optical counterpart of HLX-1 is still debated. F12  show that it is consistent either with a old ($>10$ Gyr) massive ($\sim{}2\times{}10^6$ M$_\odot{}$) SC or with a young ($\approx{}10$ Myr) SC, depending on the level of disc irradiation by the ULX.

In paper~I, we speculated that the counterpart of HLX-1 may be a disrupted satellite galaxy, undergoing a minor (1:20) merger with the S0 galaxy. According to this scenario, the BH powering HLX-1 was originally located at the centre of the satellite galaxy. In this paper, we investigate this scenario in more details, by comparing high-resolution $N-$body simulations with the available {\it HST} data.
In particular, we present a set of six simulations of the merger between a S0 galaxy and a smaller gas-rich bulgy disc galaxy. These simulations differ for the orbital properties of the satellite galaxy and for the DM content of the two galaxies.

For comparison with the simulations, we re-analyzed the {\it HST} photometric data of the HLX-1 counterpart and of the bulge of ESO~243-49, ranging from the FUV to the infrared (H) band. Firstly, we derived surface brightness profiles in five filters (F140LP, F300X, F390W, F555W and F775W) for the HLX-1 counterpart.  We find that the counterpart of HLX-1 is consistent with a point-like source in most filters. In the FUV filter, the HLX-1 counterpart shows an extended emission (Fig.~\ref{fig:fig4})  that goes beyond the PSF limits. 
 From FUV photometry alone, it is hard to distinguish which extended structures belong to the HLX-1 counterpart and which belong to the BGG or even to the disc of ESO~243-49. On the other hand, it seems unlikely that none of these structures is physically connected with HLX-1.

The lack of extended emission from the HLX-1 counterpart in the redder filters is not at odds with the merger scenario. In fact, we show (e.g. Fig.~\ref{fig:fig6} and Table~4) that the contribution of the stellar halo and of the tidal tails surrounding the nucleus of the disrupted satellite can be well below the observed surface brightness profile of the HLX-1 counterpart, even in the redder bands, provided that the merger is in a sufficiently late stage. 

The requirement that the magnitude of the simulated satellite galaxy does not exceed the observed magnitude of the HLX-1 counterpart in the redder bands (F160W and F775W) favours late stages of the merger and also the runs with the most bound orbits (run~C and run~D). In particular, run~C at $t=2.5-3$ Gyr since the last pericentre passage represents our fiducial run.

Orbits with initial relative velocity between the CMs of the two galaxies $v_{\rm rel}>100$ km s$^{-1}$  and/or with specific orbital energy $E_{\rm s}>-1.7\times{}10^4$ km$^2$ s$^{-2}$ can hardly account for the  observed magnitude of the HLX-1 counterpart in the redder bands. This implies that, if the counterpart of HLX-1 is the nucleus of a disrupted galaxy, such galaxy and ESO~243-49 must have been members of the same group of galaxies before falling into the cluster Abell~2877, as the velocity dispersion in this cluster is higher than the required $v_{\rm rel}$ to match the properties of the HLX-1 counterpart. The projected one-dimensional velocity offset between the CM of the satellite and that of the S0 galaxy in the simulated orbits is consistent with a velocity offset of $\approx{}400$ km s$^{-1}$ (Soria \&{} Hau 2012; Soria et al. 2013).

The surface brightness profiles of the HLX-1 counterpart, as well as the integrated magnitudes within $\sim{}0.4$ arcsec (beyond this value the background of the S0 galaxy dominates in the F775W and F555W filters) are in fair agreement with the contribution of the simulated stellar population in the disrupted satellite  galaxy (as derived from  run~C at $t=2.5$ Gyr after the first pericentre passage), plus a model for the irradiated disc  (Fig.~\ref{fig:fig9} and Table~5).

 The SED of the bulge of ESO~243-49, derived from the {\it HST} observations, cannot be accurately reproduced by stellar population models if a single old stellar population is assumed. 
This result is the main support for the existence of a young stellar population in the bulge of ESO~243-49.

The young stellar population may be the effect of a minor, recent SF episode,  triggered by the minor merger. In our simulations, we show that such an episode of SF is triggered by the stripping of the gas initially present in the satellite galaxy. The SF takes place mostly in the nucleus of the satellite galaxy for the first $\approx{}1$ Gyr since the first pericentre passage. At later stages, when most of the gas is stripped from the satellite galaxy and is accreted by the primary galaxy, almost all the SF takes place in the bulge of the S0 galaxy. From the simulations, we see that the SF in the bulge of the S0 goes on quietly for several Gyr (even after the end of the merger), at a level of $\sim{}10^{-3}-10^{-2}$ M$_\odot{}$ yr$^{-1}$ (Fig.~\ref{fig:fig10}).

 We derive the expected SED for the bulge of the S0 galaxy directly from our run~C and we find that it reproduces the observed data quite well. The main difference between the SED obtained from run~C and the observed SED is in the FUV, where the simulation accounts only for $\sim{}60$ per cent of the observed flux.

 In summary, the comparison between the {\it HST} photometric data of ESO~243-49 and the results of our simulations confirms that a minor merger is a viable scenario to explain the properties of HLX-1 and of its counterpart. 
 The nature of the extended FUV emission surrounding the counterpart of HLX-1 deserves further investigation. 
On the other hand, we suggest that  definitive evidence for a recent minor merger between ESO~243-49 and the counterpart of HLX-1 might come from new kinematic measurements.

\section*{Acknowledgments}
We thank the anonymous referee for the critical reading of the manuscript and for the comments that improved it significantly. We thank the authors of gasoline (especially J. Wadsley, T. Quinn and J. Stadel), L.~Widrow for providing us the code to generate the initial conditions, L.~Girardi and A.~Bressan for providing the SSP models. We also thank E. Ripamonti, E. Bertone, L. R. Bedin and L. Mayer for useful discussions. To analyze simulation outputs, we made use of the software TIPSY\footnote{\tt http://www-hpcc.astro.washington.edu/tools/tipsy/\\tipsy.html}.
The simulations were performed with the {\it lagrange} cluster at CILEA and with the PLX cluster at the CINECA. We acknowledge the CINECA Award N. HP10CLI3BX and HP10B3BJEW, 2011 for the availability of high performance computing resources and support. MM and LZ acknowledge financial support from INAF through grant PRIN-2011-1. MM acknowledges financial support from the Italian Ministry of Education, University and Research (MIUR) through grant FIRB 2012. LZ acknowledges financial support from ASI/INAF grant no. I/009/10/0. We thank the NASA/IPAC Extragalactic Database (NED).


\end{document}